\documentclass[journal]{IEEEtran}
\IEEEoverridecommandlockouts
\usepackage{graphicx} 
\usepackage{tabularx}
\usepackage{ulem}
\usepackage[T1]{fontenc}
\usepackage{algorithmic}
\usepackage{comment}
\usepackage{amsmath,amssymb,amsfonts}
\usepackage{array}
\usepackage{textcomp}
\usepackage[font=small]{caption}
\usepackage{paralist}
\usepackage{lettrine}
\usepackage{amsthm}
\usepackage{cite}
\usepackage{xcolor}
\usepackage{cancel}
\usepackage{booktabs}
\usepackage{subcaption}

\usepackage{stfloats}
\usepackage{url}
\usepackage{lipsum}
\usepackage{enumitem}
\usepackage{mathtools}
\usepackage{cuted}
\usepackage{makecell}
\usepackage{acronym}
\usepackage{multicol}
\usepackage{multirow}
\usepackage[ruled,vlined]{algorithm2e}
\usepackage{authblk}
\usepackage{fixltx2e}
\usepackage{tikz}
\usepackage{hyperref}

\usepackage[utf8]{inputenc}
\usepackage{cite}
\usepackage{floatrow}

\usepackage{amsmath,amssymb,amsfonts,stfloats}
\usepackage{tabularx}
\usepackage{algorithmic}
\usepackage{graphicx}
\usepackage{textcomp}
\usepackage{floatrow}
\usepackage{xcolor}
\usepackage{booktabs}

\usepackage{multicol}
\usepackage{acronym}
\usepackage{url}
\usepackage{mathtools}
\usepackage[font=small,labelfont=bf]{caption}
\graphicspath{{./}} 

\begin{document}

\title{
Mobile Radio Networks and Weather Radars Dualism: Rainfall Measurement Revolution in  Densely Populated Areas
\thanks{D. Tornielli Bellini, D. Tagliaferri and U. Spagnolini are with the Department of Electronics, Information and Bioengineering, Politecnico di Milano, Via Ponzio 24/5, 20133 Milano, Italy.  e-mail: \{davide.tornielli,dario.tagliaferri,umberto.spagnolini\}@polimi.it}
\thanks{S. Duque, is with Huawei Technologies, Munich Research Center,  Riesstraße 25, 80992 Munich, Germany.  e-mail: sergio.duque.biarge@huawei.com}
\thanks{L. Resteghini, is with Milan Research Center, Huawei Technologies Italia S.r.l., Milan, Italy 
e-mail: {laura.resteghini@huawei.com}}
\thanks{L. Baldini and E. Adirosi are with National Research Council of Italy, Institute of Atmospheric Science and Climate (CNR-ISAC), via Fosso del cavaliere, 100, 00133, Rome, Italy. e-mail: \{luca.baldini,elisa.adirosi\}@cnr.it}
\thanks{M. Montopoli  (corresponding Author) is with
the National Research Council of Italy, Institute of Atmospheric Science and Climate (CNR-ISAC), via Fosso del cavaliere, 100, 00133, Rome, Italy, and with Center of Excellence for Telesensing of Environment and Model Prediction of Severe events, Università Dell’Aquila, L’Aquila, 67100 L’Aquila, Italy  e-mail: mario.montopoli@cnr.it}
}

\author{Davide~Tornielli~Bellini,~\IEEEmembership{Graduate Student Member,~IEEE,}  
Mario~Montopoli, 
Dario~Tagliaferri,~\IEEEmembership{Member,~IEEE,} 
Luca~Baldini,~\IEEEmembership{Senior Member,~IEEE,}
Elisa~Adirosi, 
Sergi~Duque, 
Laura~Resteghini,
Umberto~Spagnolini,~\IEEEmembership{Senior Member,~IEEE}
}

\maketitle

\begin{abstract}

This study demonstrates, for the first time, how a network of cellular base stations (BSs)—the infrastructure of mobile radio networks—can be used as a distributed opportunistic radar for rainfall remote sensing. By adapting signal‑processing techniques traditionally employed in Doppler weather radar systems, we demonstrate that BS signals can be used to retrieve typical weather radar products, including reflectivity factor, mean Doppler velocity, and spectral width.
Due to the high spatial density of BS infrastructure in urban environments, combined with intrinsic technical features such as electronically steerable antenna arrays and wide receiver bandwidths, the proposed approach achieves unprecedented spatial and temporal resolutions, on the order of a few meters and several tens of seconds, respectively. Despite limitations related to low transmitted power, limited antenna gain, and other system constraints, a major challenge arises from ground clutter contamination, which is exacerbated by the nearly horizontal orientation of BS antenna beams. This work provides a thorough assessment of clutter impact and demonstrates that, through appropriate processing, the resulting clutter-filtered radar moments reach a satisfactory level of quality when compared with raw observations and with measurements from independent BSs with overlapped field-of-views. The findings highlight a transformative opportunity for urban hydrometeorology: leveraging existing telecommunications infrastructure to obtain rainfall information with a level of spatial granularity and temporal immediacy like never before.
 
\end{abstract}


\section{Introduction}

\IEEEPARstart{Q}{uantitative}  precipitation estimation (QPE) underpins hydrological modeling, flood and landslide early warning systems, drought assessment, and climate diagnostics. However, the pronounced spatial and temporal variability of rain fields makes accurate observation intrinsically challenging. Traditional surface networks of rain gauges provide direct measurements of accumulated rainfall but suffer from sparse, uneven global distribution and point-scale representativeness, especially over complex terrain.
Optimal rain gauge placement is not always applicable due to logistic issues and integration of  gauges with spatially continuous precipitation estimates often  carried out \cite{suri:2024, vieux:2005, silvera:2019}. 
Weather radars offer over-land high-resolution spatial snapshots of precipitation at minutes time scale, but they infer surface rain rates through microphysical assumptions and algorithms that are sensitive to environmental and instrumental effects \cite{song:2017, silvera:2019, ryu:2025}. Satellite remote sensing fills large observational gaps by providing quasi-global coverage and near–real-time precipitation maps. However, satellite retrievals remain uncertain for light/shallow precipitation, in particular orographic contexts, and mixed-phase hydrometeors, and typically require ground validation and bias correction. These complementary strengths and weaknesses motivate integrated frameworks that merge satellite, radar, and gauge information to improve accuracy and spatiotemporal completeness in QPE for both science and applications \cite{skofronick:2018, belay:2022}.

Recognizing the trade-offs among methods, integrated precipitation estimation merging gauges, radar, and satellite has become a central line of research. Emerging non-traditional opportunistic  observations, like those from commercial microwave links (ground to ground or ground to satellite) , citizen science, vehicle sensors, and camera-based techniques can complement established systems, but require rigorous quality control before operational adoption \cite{jhm:2023, belay:2022}.

\begin{figure}[!t]
    \centering
     \includegraphics[width=0.95\columnwidth]{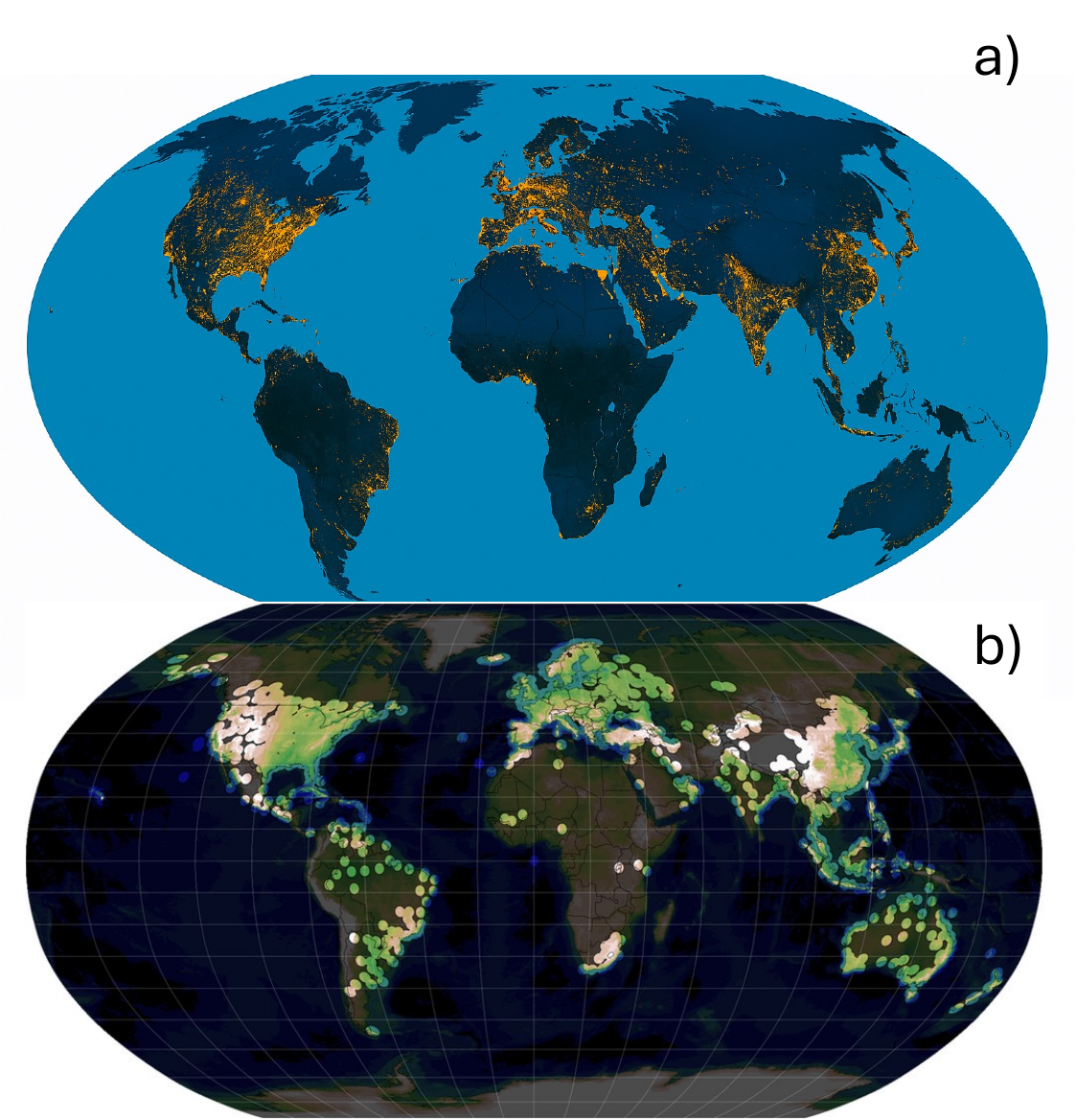}
  \caption{Conceptual visualization of BS distribution reflecting the global coverage trends derived from publicly available sources and industry reports \cite{GSMA:2025,OpenCelliD:2025,ITU:2025,Ookla:2025}. Yellow/Orange areas indicate high density (urban regions in North America, Europe, East Asia). Dark green areas show moderate coverage. Sparse regions represent limited connectivity a); Weather radar coverage at 2019 (with permission from \cite{Saltikoff:2019b}) in b).}
 
  \label{fig:WR_and_BS_distributionWorldWide}
\end{figure}

Recent advances have envisioned the exploitation of wireless networks endowed with integrated sensing capabilities to extract new source of information from the surrounding environment \cite{Nuria:2024, Mantuano:2025, Polisano:2025, Beni:2026}. The underlying principle is to leverage these networks, primary designed for interconnecting end-user equipments (e.g., mobile devices), to deliver services that extend beyond conventional communication. This can be achieved incorporating the base stations (BS), i.e. the network nodes responsible for routing information to end-user devices such as mobile phones, in a new service able to  generate a novel data stream that bridges environmental geophysical variables with the radio-frequency domain. 
Following this new paradigm, a very preliminary conceptualization of the use of BSs for precipitation sensing was introduced in \cite{11173435}, focusing on antenna beamforming design to maximize the spatial rejection of ground clutter. However, the authors in \cite{11173435} employ a BS  prototype designed for short-range operation and targeted toward 6G applications, whereas in this work a commercial 5G-Advanced BS is utilized, which supports communication over distances of several kilometers and corresponds to a more widely deployed and standardized platform. In addition, the study in \cite{11173435} does not address the practical challenges associated with integrating precipitation sensing into the standard operation of a BS. These challenges include the analysis of the limited sensitivity of BS system to precipitation, the use of a physically based rain scattering models to describe precipitation fields over the resolution volumes and experimental evidences using actual measurements acquired in both clear sky clutter-only situations as well as during rain precipitation events.
Conversely, this study demonstrates, for the first time, with the aid of physically-based simulations, the feasibility of using BS in weather radar-like mode (BS-WRM) to track extreme precipitation, opening to a new way to look at the opportunistic radio signal observations.
 The proposed approach leverages the radar-like sensing capabilities of BS to monitor rainfall, and, owing to the extensive global deployment of BS infrastructure (Figure \ref{fig:WR_and_BS_distributionWorldWide}a), particularly in densely populated areas, this technique exhibits remarkable scalability and considerable operational attractiveness. If compared with the weather radar coverage worldwide in figure \ref{fig:WR_and_BS_distributionWorldWide}b), it can be seen as BS distribution closely follows that of weather radars but with some important differences related to approx 10-times higher spatial resolution and refresh time offered by BS than custom weather radars.
The integration of BS data into QPE frameworks holds significant promise for enhancing multi-source data fusion alongside conventional ground-based radars, satellite sensors, and rain gauge networks.  
It is important to note that, unlike the commercial microwave link approach, which relies on bistatic radio connections that leverages the path-integrated signal's attenuation to infer an average precipitation intensity, the BS based methodology directly leverages end-user communication links in a radar-like monostatic scheme, thus enabling range-resolved inference of the precipitation field. 
The pervasive capillarity of BS compared with any other  opportunistic sensing tool, makes the themes treated relevant for improve precipitation field mapping especially in urban areas where there exist an observational gap of precipitation measurements.

The paper is organized into seven sections. Section II provides a review of precipitation remote sensing at microwave frequencies, in order to contextualize the proof of concept for using BS as a novel source of local rainfall information. Section III presents an overview of BS characteristics, whereas Section IV describes the methodology adopted for processing the BS data. Section V demonstrates rainfall detection from BS using simulated data, while Section VI is devoted to discussing the implications of the proposed methodology. Finally, Section VII presents the conclusions.

\section{State of the art on remote sensing of precipitation}

This section covers a concise overview of the state-of-the-art techniques used to monitor atmospheric precipitation worldwide, using sensors working in the microwave frequency bands. This is particularly useful for establishing a framework that enables a better interpretation of observations based on the BS approach proposed in this work.

\subsection{Satellite active and passive systems}
\label{sect:Sat}
Satellite-based active and passive microwave sensors deliver complementary insights into global precipitation, distinguished by their spatial resolution and temporal coverage.

Active sensors onboard low-Earth orbit satellites capture vertical profiles via dual-frequency Ku (13.6GHz)-/Ka(35.5GHz)-band radars, resolving hydrometeor vertical structure at $\sim$ 5 km $\times$ 250 m sampling, revisiting the same region every few hours when integrated into an active-passive constellation framework\cite{skofronick:2018}. 
 As demonstrated by the NASA/JAXA Global Precipitation Measuring Mission (GPM) \cite{skofronick:2018} and the Chinese  Feng Yun 3G (FY-3G) \cite{Zhou:2025} mission,  dual-frequency techniques and path integrated attenuation constraints, yield improvements in rain rate and drop size distribution (DSD) retrieval, including light rain and snowfall, although non-uniform beam filling and complexities in the melting layer remain challenging \cite{Grecu:2020}.

Passive radiometers operating at 10–183 GHz offer wide-swath coverage with footprints ranging from $\sim$15 km at high frequencies to $\sim$50 km at low frequencies. Bayesian retrieval frameworks (e.g. GPROF algorithm \cite{Pfreundschuh:2024, ryu:2025} and its neural network version: GPROF-NN \cite{Pfreundschuh:2022}) have reached a good degree of maturity providing rain-rate estimates over spatial scales of $\sim$15–30km.

Active and passive synergy are often used to align high resolution vertical structure from radar with the broader spatial sampling of radiometers. These methods help alleviate sampling mismatches and improve retrieval consistency \cite{Grecu:2020, Rahimi:2025}. Global products like IMERG consolidate the multi-sensor constellation into $\sim$10 km spatial grids at 30-minute intervals, enabling near-real-time hydrological and climatological applications \cite{Duque:2023, Huffman:2024}.


\subsection{Opportunistic radio signals}
\label{sect:CMLs}

Opportunistic sensing exploits existing telecommunication infrastructures to infer precipitation from rain‑induced attenuation of microwave signals. Two main link types are used: Commercial Microwave Links (CMLs) in terrestrial cellular backhaul networks and Satellite Microwave Links (SMLs) such as TV‑SAT or broadband terminals. Their characteristic spatial and temporal resolutions are summarized in Table \ref{tab:cml_sml_resolution}. Operating above 6GHz, both systems experience scattering and absorption by raindrops, making path‑integrated attenuation a proxy for integrated rain rate. As a consequence, neither CMLs nor SMLs resolve the vertical precipitation structure, and geolocating rainfall becomes increasingly uncertain for long link paths.
CML‑based rainfall estimation has advanced substantially, enabling high‑resolution urban monitoring and supporting flood‑forecasting applications \cite{Cazzaniga:2022, Liu:2023, Zhang:2023b, Nature:2023, Janco:2023, Nebuloni:2025b}. SMLs extend this concept to slanted paths, providing broader spatial coverage and higher temporal sampling \cite{Biscarini:2020, Adirosi:2020, Scognamiglio:2024, Angeloni:2024, Nebuloni:2025}. Despite these developments, important challenges persist: accurate attenuation‑to‑rain conversion requires reliable atmospheric emission models and rain‑height information; wet‑antenna effects;  non‑uniform rain fields along the path bias; and link heterogeneity complicates integrated processing. Furthermore, link density remains limited in many regions (Table\ref{tab:link_density}), reducing spatial sampling and hindering homogeneous monitoring.
Future prospects include fusion of CML and SML observations, physics‑informed machine‑learning approaches \cite{Kumar:2024}, and standardized open datasets to transform communication networks into dense environmental sensing systems for hydrology and climate resilience \cite{Graf:2026, Saggese:2022}. Although fiber‑optic connections are expected to replace microwave backhaul on major network backbones, microwave links will remain essential where cabling is impractical (e.g., rural areas, wetlands), implying that CML availability in urban settings will gradually decline.

\begin{table*}[t]
\centering
\caption{Indicative spatial and temporal resolution and operating frequency bands for opportunistic precipitation sensing using Commercial Microwave Links (CMLs) and Satellite Microwave Links (SMLs). Ranges are representative of recent deployments; actual values depend on network geometry, sampling strategy, and processing.}
\label{tab:cml_sml_resolution}
\begin{tabular}{l|ccc}
\hline
\textbf{Link type} & \textbf{Frequency band (GHz)} & \textbf{Path scale (km)$^{\,a}$} & \textbf{Native sampling (s)$^{\,c}$} \\
\hline
CML                     & 6--40              & \(\sim 0.2\)--10    & 1--10 (research);  \\[2pt]
(terrestrial backhaul)  & (typ. 18, 23, 38)  & (typ. 1--4)        & 60--900 (operator archives) \\[2pt]

SML                               & Ku: 10.7--14; & \(\sim 2\)--10      &    10--60 (typical telemetry); \\ 
(TV\,-SAT / downlinks)& Ka: 18--30    & slant path segment   &  up to\(\sim\)300 \\ 
\hline
\end{tabular}

\vspace{6pt}
\begin{minipage}{0.95\linewidth}\footnotesize
\textit{Notes.}
$^{a}$ Length of the path (or effective rainy segment) over which path attenuation is measured.\\
$^{b}$ Native time sampling of received signal level (or SNR) available from operators/terminals; research systems can sample faster than operational archives.
\end{minipage}
\end{table*}

\begin{table}[!]
\centering
\caption{Indicative density ranges of Commercial Microwave Links (CML) and Satellite Microwave Links (SML) for opportunistic precipitation sensing, expressed as links per 100\,km$^2$. Values are conceptual estimates based on published case studies \cite{RiosGaona2018,Graf2020,Graf:2026}, ITU reports (\cite{ITU2025_GMDI_SecondRoundTable,ITU2024_GMDI_RoundTable,ITU2016_L1700_Sup23}), and regional adoption patterns.}
\label{tab:link_density}
\begin{tabular}{lcc}
\hline
\textbf{Region / Corridor} & \textbf{CML Density} & \textbf{SML Density} \\
                           & (links/100\,km$^2$) &  (links/100\,km$^2$) \\
\hline
Western Europe             & 15--25  & 2--8  \\
Eastern North America      & 10--20  & 3--10 \\
Western North America      & 6--12   & 2--6  \\
East Asia                  & 12--22 & 2--6  \\
South Asia (India)         & 8--15   & 1--4  \\
Southeast Asia             & 10--18  & 1--3  \\
SE Brazil                  & 8--14   & 1--3  \\
Southern Africa            & 5--10   & 0.5--2\\
Middle East                & 6--12   & 1--3  \\
SE Australia               & 5--10   & 1--3  \\
\hline
\end{tabular}
\end{table}

\subsection{Ground based weather radar systems}

Ground-based weather radars provide high-resolution volumetric observations of precipitation, typically with radial resolutions of 100m to 1km and angular resolution of 1°, usually resampled, for applications, onto a  Cartesian grids at 0.5–1km. Temporal updates are frequent, with operational weather radar networks such as NEXRAD \cite{Fulton:1998, Weber:2021},  OPERA \cite{Huuskonen:2014,Saltikoff:2019} and CINRAD \cite{Zhao:2019, Shao:2025} in US, Europe and China, respectively, delivering 5-minute refresh cycles. These characteristics make radars indispensable for monitoring convective systems and short lived precipitation extremes. Obviously, weather radars are installed exclusively over land, which inherently restricts their observational coverage to terrestrial areas and portions of ocean very close to the coast. Typical frequencies allocated for weather radars are within S-band ($\sim$ 2.7GHz), C-band ($\sim$ 5.6GHz) and X-band ($\sim$ 9.4GHz) (see table \ref{tab:radar_specs}). 
The country or continental-wide networks that are mostly built using S-band and/or C-band radars, have been designed with the aim of maximizing coverage while keeping costs low, being costs driven primarily by the number of radars installed. These networks are typically configured to minimize overlap between individual radar coverage areas. Unavoidably, coverage gaps can occur due to Earth curvature, beam overshooting beyond $150$km, and beam blockage due to complex terrain or human-made structures.
To fill those gaps, X-band weather radars have emerged as a critical component for enhancing precipitation monitoring in critical areas.  X-band weather radar local networks emerged with the CASA (Collaborative Adaptive Sensing of the Atmosphere) program in the United States \cite{Wang:2010} and found application in large urban regions like in the Dallas–Fort Worth (DFW) urban testbed \cite{Brewster:2017}, and Tokyo Metropolitan Area (Japan) \cite{Asai:2021}.
In the latter case, phased array radar antennas are used. Although conventional weather radar are dominated by system with mechanical moving antennas (i.e. mechanically steered parabolic dish), systems equipped with phased arrays attempt to emerge \cite{Kollias:2018,USHIO:2024}. Phased array radars (PAR) utilize electronically steered antenna arrays which allow eliminating mechanical inertia. This  enables rapid and adaptive single or multi-beam scanning,  with full-volume updates achievable in tens of seconds and targeted sector scans in sub-second intervals. Such capability significantly enhances temporal resolution for monitoring rapidly evolving convective storms and urban flash-flood hazards.  However, PAR systems entail initial higher costs and complexity, require sophisticated calibration to manage beam shape and sidelobes suppression, and remain less widely deployed than conventional networks. Dual polarization capability is also offering a supplementary advantage  over single polarization systems in terms of data quality (e.g. more accurate clutter removal,  attenuation effects compensation) and refined quantitative precipitation estimates \cite{Bringi:2001, zhang:2016,Montopoli:2017}. However, such improvements  are conditioned to  the availability of low noise polarimetric variables.  Challenges remain in cost, maintenance, and data fusion, but operational benefits in disaster risk reduction and urban resilience underscore the value of radar networks.

\begin{table*}[htbp]
\centering
\caption{Typical system features of weather radars at S-, C-, and X-bands. Bold numbers in the round brackets refers to values used in the simulations of figure \ref{fig:radar_sim}.}
\begin{tabular}{l|c|c c|c c|c c}
\hline
\textbf{Parameter} & \textbf{Symbols} &\textbf{S-Band}&  &\textbf{C-Band} & &\textbf{X-Band} \\
\hline
Frequency (GHz)             & $f_0$            & (2.7, 3.0)  & (\textbf{2.7}) & (5.4, 5.8) & (\textbf{5.6})  & (9.3, 9.6)    & (\textbf{9.4})      \\
Peak Input Power (kW)       & $P_{\rm tx}$          & (750, 1000) & (\textbf{750}) & (200, 350) & (\textbf{200})      & (25, 100) & (\textbf{100})        \\
Noise Figure (dB)           & $F$           &(3, 5)       & (\textbf{4})   & (3, 5)     & (\textbf{4})     & (3, 5)     & (\textbf{4})      \\
Pulse Length ($\mu$s)       & $\tau_{\rm tx}$         & (0.3, 2.0)  & (\textbf{0.33})& (0.3, 2.0) & (\textbf{0.33})      & (0.3, 2.0)& (\textbf{0.33})      \\
Antenna Max Gain (dBi)       & $G_{\rm max}$      & (45, 46)    & (\textbf{43})   & (44, 45)  & (\textbf{43})       & (43, 44) & (\textbf{43})       \\
HPBW Azimuth ($^\circ$)     & $\Delta \phi$   & $\approx$1.0& (\textbf{1})  & $\approx$1.0 & (\textbf{1})     & (1.0, 1.5)& (\textbf{1})      \\
HPBW Elevation ($^\circ$)   & $\Delta \theta$ & $\approx$1.0& (\textbf{1})   & $\approx$1.0& (\textbf{1})     & (1.0, 1.5)& (\textbf{1})    \\
Polarization Scheme         & -              & Dual-pol (H/V)& &Dual-pol (H/V)&   & Dual-pol (H/V)&  \\
\hline
\end{tabular}
\label{tab:radar_specs}
\end{table*}

\section{Base station system}
\label{sect:BSsysetm}

\begin{figure}[!t]
    \centering
     \includegraphics[width=0.9\columnwidth]{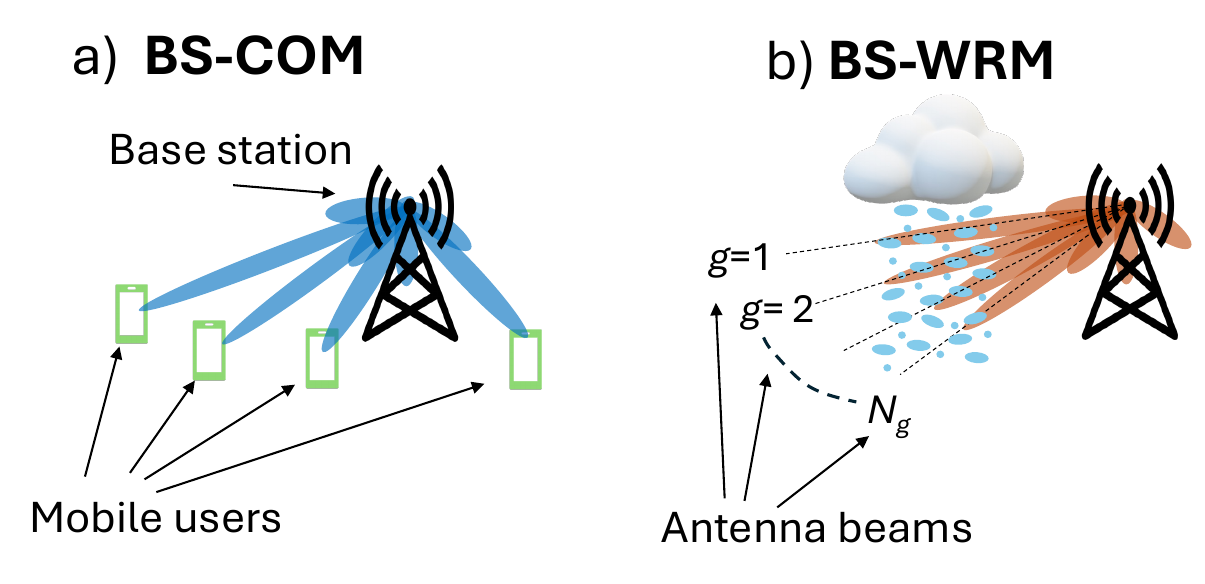}
  \caption{
  Conceptual figure of a base station (BS) in a typical communication mode (COM) in which BS serves the final mobile users (mobile devices in green) a); and the weather radar mode (WRM) in which BS antenna multi beams (orange lobes) scans a sectorial portion of the cell to intercept rain in the area covered b).}
  \label{fig:ConceptualBS}
  \end{figure}

This section describes the usage of base stations of cellular networks as opportunistic weather radars. In cellular communication systems, a BS provides wireless connectivity between users and the core (wired) communication network. Each BS is responsible for transmitting and receiving radio-frequency signals within a defined coverage area, commonly referred to as a cell. Hereafter, we will refer to the two BS operating modes as communication standard mode (COM), which is the primary BS functionality (figure \ref{fig:ConceptualBS}a), and an experimental weather radar mode (WRM), which is the application discussed in the present work \ref{fig:ConceptualBS}b). \\

\subsection{Spectrum allocation} In typical BS-COM operation (of the 5G or next-generation wireless networks) each group of BSs pertaining to the same operator is allocated with a pre-defined spectrum portion, centered around carrier frequency $f_0$ and of fixed bandwidth $B$. In current 5G standard, regulated by the third generation partnership project (3GPP) and the international telecommunication union (ITU), the allowed carrier frequency $f_0$ is within the so-called frequency range 1 (FR1, $f_0 < 6$ GHz) \cite{itur_m1036} while next-generation of wireless networks (6G) will likely extend the operation to frequency range 3 (FR3, $7 \leq f_0 < 24$ GHz). The typical available bandwidth at FR1, which is the operating frequency considered in this work, amounts to $B=100$ MHz, while FR3 promises to guarantee much wider bandwidths \cite{itu_r_m2160}. On the same bandwidth, each BS is responsible of a different cell, minimizing the mutual interference and ensuring coexistence by suitable spatial reuse of the spectrum \cite{rappaport_wireless_2023}.

\subsection{Antenna configuration} 
The single BS is typically equipped with three antenna arrays, each covering an azimuthal sector of $120^\circ$. Each antenna is made by a uniform rectangular array with $N_\phi$, $N_\theta$ patch antennas along the azimuth and elevation, respectively. The spacing among the patches is optimized  to improve the beamwidth for communication purposes, conventionally using a slanted ($\pm 45$ deg) polarization. Such arrays have full electronic scanning capabilities and can implement flexible beampatterns according to needs. In typical operation, the BS-COM implements \textit{beamforming}\footnote{The beamforming is either digital or analog, mainly depending on the carrier frequency. Digital beamforming allows flexible beam-pattern generation at the expense of a higher hardware and processing cost, especially for analog-to-digital and digital-to-analog components, and it is therefore widespread only at carrier frequencies $f_0<6$ GHz.} to maximize the data-rate to/from the users by directional transmission/reception in multiple directions. The minimum half power  beamwidths (HPBWs) along azimuth ($\Delta \phi$) and elevation ($\Delta \theta$), is ruled by $N_\phi$ and $N_\theta$ respectively, with  typical array footprints of the order of few degrees and maximum antenna gains that range from $20$ to $30$ dBi.
  
\subsection{Waveform, frame organization and emitted power} 
\label{sect:FrameOrg}
\begin{figure*}[!b]
    \centering
     \includegraphics[width=1\columnwidth]{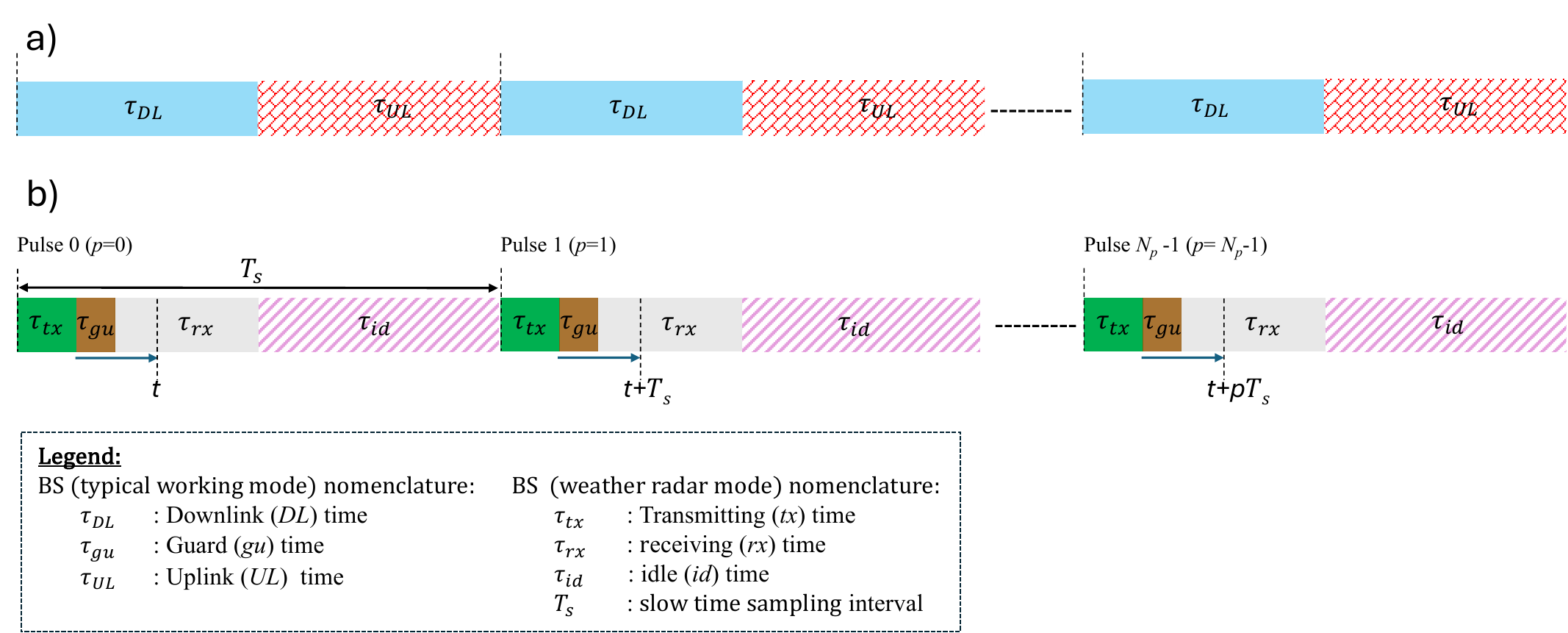}
  \caption{Representation of time frame organization a) in the typical BS-COM working mode; b) in the experimental BS-WRM. Note that in panel a), downlink (DL) and uplink (UL) time slots are depicted as separate  and contiguous slots for ease of illustration, whereas in practice they  interleave using a more articulated pattern following the 3GPP standard.}
  \label{fig:BSframe}
  \end{figure*}
  
The 3GPP standard-compliant BSs operate in \textit{half-duplex}, alternating downlink (DL) and uplink (UL) communication phases in which the BS is transmitting (serving the users) or receiving information from the users, respectively, in a  \textit{time division duplexing} working mode. The DL and UL time slot, denoted as $\tau_{DL}$ and $\tau_{UL}$, are typically interleaved according to predefined patterns in order to operate BS-COM under its nominal operating conditions (see Figure~\ref{fig:BSframe}a).
The maximum continuous power emitted by the BS during DL is limited by regulations, implying an emitted power $P_{\rm tx}$ of tens to hundreds of Watts depending on the antenna gains.
In both periods (DL and UL), the BS employs orthogonal frequency division multiplexing (OFDM). In OFDM, the available bandwidth, $B$, is divided in $N_{\rm sub}$ subcarriers, each of them at frequency $f_n = f_0 + n\delta f$, $n=0,...,N_{\rm sub}-1$, where $\delta f$ is the subcarrier spacing, chosen in a pre-defined pool according to the 3GPP numerology \cite{3gpp38211_v18}. To simplify, the transmission is organized in symbols, each of duration $1/\delta f$ so that to allow the orthogonality of the subcarriers
and symbols are grouped in frames, whose duration is a flexible parameter that can be optimized by the BS according to the specific context, constrained by the required alternation between DL and UL operation, which tends to follow a regular and periodic or cyclic pattern over time (see figure \ref{fig:BSframe}).

\subsection{Differences and Similarities Between a BS and a Weather Radar }
The BS technology strikes an interesting parallelism between a radar-like BS, termed as BS-WRM, and specifically designed  weather radars.
Differences between the two systems interest antenna technology, transmission hardware and the acquisition schedule.   

\subsubsection{Antenna scan mode and coverage strategies} Most of weather radars make use of mechanical scans except in few cases that uses phased array technology. The latter is more closely related to that performed by a typical BS-COM that uses arrays of  patch antennas to electronically scan the surrounding scene. 
Aside from the differences in antenna performance (BS-COM has nearly half of the antenna gain of a weather radar with a much wider HPBW up to 6 times higher), another important difference to underline is related to the fact that a weather radar is optimized to detect the surrounding environment indiscriminately by performing volumetric scans with at given repetition cycle, whereas the BS-COM is designed to guarantee the capacity coverage within the cell, thus in this case multi-beams are directed towards users, and no full scanning is implemented, except for quasi-periodic idle periods in which a BS scans the coverage area to search for new users. 
\subsubsection{Antenna elevation angles}
In terms of siting,  BSs are typically located at a given height from ground (ranging from $5-6$ m to $20-30$ m) and tilted downward with elevation angle with respect to horizon $\theta <0^\circ$, in order to maximize the illumination of the users. In such circumstance, the illuminated scene is dominated by the ground clutter (e.g. buildings, civil infrastructures, bridges, etc.). Therefore, operate a BS as a weather radar would require the implementation of dedicated de-cluttering methods (see later sections). 
It is worth noting that, although the BS antenna is typically downtilted in order to comply with local regulatory requirements and to optimize data transmission performance, dedicated configurations in which the beams are directed upwards are technically feasible and cannot be ruled out for future implementations.

\subsubsection{Acquisition cycle} As described in section \ref{sect:FrameOrg}, BS-COM typically operates a pattern scheme of  DL transmission and UL reception time slots for   $\tau_{\rm DL}$ and $\tau_{\rm UL}$, respectively (figure \ref{fig:BSframe}a). 
Such a scheme is typical for a BS in its native COM mode but it can  be easily adapted to a typical pulsed WRM. This can be achieved by allocating a time period $\tau_{\rm tx}$ for transmission and reserving some of the remaining time ($\tau_{\rm rx}$) for the reception of the back-scattered radar echoes (figure \ref{fig:BSframe}b). However, since both COM and WRM must coexist together, only a small fraction ($\tau_{\rm rx}$) of the total available time ($T_s$) is dedicated to the reception of the meteorological signals. The time $\tau_{\rm gu}$ is a sort of guard period  that guarantees a clear separation between  the transmitted signal and the received one making the latter more easily detectable. As a consequence, a blind zone caused by  $\tau_{\rm tx}+\tau_{\rm gu}$ arises. However, such blind zone is expected to be of the order of the Inter-Site Distance (ISD). Therefore, the blind zone of one BS could be likely covered by other BSs in the surrounding area.
The rest of the available intra-frame time ($\tau_{\rm id}$) is an idle period from the point of view of the WRM, whereas it is an active UL time for the COM. Indeed, during $\tau_{\rm id}$ the BS could continue to operate a DL and UL sequence thus continuing to guarantee the COM service. 
Having figure \ref{fig:BSframe}b in mind, one of the main difference in the acquisition cycle of  BS-WRM, with respect to more customary pulsed weather radar systems, stems on the fact that the acquisition of meteo-signals is not continuous but intermittent in BS-WRM being  the sequences of $\tau_{\rm tx}$ and $\tau_{\rm rx}$ interleaved by an idle period $\tau_{\rm id}$.
The selection of $\tau_{\rm tx}$, $\tau_{\rm rx}$ and $\tau_{\rm id}$, is quite flexible but it must comply, in the first instance, to ensure the capacity of the BS-COM service through an adequate time $\tau_{\rm id}$. Secondary, from the standpoint of BS-WRM,  $\tau_{\rm tx}$ and $\tau_{\rm rx}$ should be properly selected to allow a sufficient signal-to-noise-ratio (SNR) and coverage in range.  Too short $\tau_{\rm tx}$ implies lower transmitted energy and consequently lower SNR. On the contrary, too large $\tau_{\rm tx}$ can erode $\tau_{\rm rx}$ restricting the maximum unambiguous radar range. However, thanks to the OFDM mode, the transmitted signal in $\tau_{\rm tx}$ has a bandwidth $B$ much larger than $1/\tau_{\rm tx}$ resulting in a pulse compression gain $B \tau_{\rm tx} \gg 1$ which allow to maintain the range resolution high in the presence of longer $\tau_{\rm tx}$ \cite{10264814} (see Section \ref{sec:processing} for details). For reasons of confidentiality, complete typical values of BS-WRM cannot be made publicity available. For what of interest here, order of magnitude of $B$ is around 20 MHz achieving range resolution around 7 m with a slow time sampling interval $T_s=\tau_{\rm tx}+\tau_{\rm gu}+\tau_{\rm rx}+\tau_{\rm id}= 2.5$ ms.

\subsubsection{Transmitted power} One main drawback, of BS-WRM is due to the low power engaged compared to weather radars. For obvious reasons,
there is significantly less available peak power (order of $10^{-3} $) in a BS-WRM than in a traditional weather radar, 
and consequently the ability to detect meteo target at far ranges will be limited (see section \ref{sect:ResultsSimScenarios}).

\subsubsection{Polarization scheme}
BS uses slanted linear  $45^\circ$ slanted linear polarizations both in the transmission and reception side. Although slanted $\pm 45^\circ$ linear polarizations is quite common during transmission in most of commercial weather radar systems, the horizontal symmetry of some meteorological hydrometeors (e.g. liquid drops) and the absence of the horizontal and vertical components in reception makes the use of polarimetry tricky for BS-WRM.

\subsubsection{Carrier frequency} The BS working frequency can  be  approximately 5.0 GHz, which is in close proximity to the 5.6 GHz band allocated for C-band weather radar services. It should be noted that the BS operating frequency is not standardized globally and may differ across regions. Nonetheless, for the purposes of simulation and comparative analysis performed, the BS operating frequency is fixed at 4.9 GHz in this study.
Although at first glance 4.9 GHz and 5.6 GHz frequency difference appears to be not significant, it has some consequences in the BS-WRM derived quantities (see section \ref{sect:ResultsSimScenarios}).

\section{Processing chain}\label{sec:processing}

The processing chain stems from the reception of I-Q signal of the BS-WRM. Let us assume that the BS-WRM reserves a periodic time interval in the standard-compliant frame organization in which it implements an active beam scanning over a predefined azimuth sector.
 We denote with $N_g$ the number of beams that the BS-WRM implements, with a single beam pattern denoted by $f_g(\theta,\phi; \theta_g,\phi_g)$ in which ($\theta_g,\phi_g$) indicates the pointing direction of the $g$-th beam, $g=1,...,N_g$. Hereafter, for sake of clarity we use the subscript $g$ to denote a quantity referred to the $g$-th beam (e.g., $f_g(\theta,\phi; \theta_g,\phi_g)$ = $f_g(\theta,\phi)$). Following the frame organization in figure \ref{fig:BSframe}b, for each beam the BS-WRM transmits a regular train of pulses identified by the pulse index $p$, whereby a single pulse comprises the effective duration, $\tau_{\rm tx}$, the switching time from Tx to Rx, $\tau_{\rm sw}$, a receiving interval, $\tau_{\rm rx}$, that enables the BS-WRM to gather the precipitation echoes and an idle interval, $\tau_{\rm id}$, which  is needed to separate weather and telecommunication services (see figure \ref{fig:BSframe}b).
 Consequently, having $N_p^{'}$ pulses per beam, the total duration of the beam sweeping is $T_{\rm tot} = N_g N_p^{'} T_s$. $T_{\rm tot}$ is periodically repeated over fairly long duty cycles (tens of seconds) to monitor and track precipitation.   

\subsection{Pre-processing} \label{subsect:preprocessing}
Let us define the Rx complex I-Q signal by the BS-WRM on the $g$-th beam and $p$-th pulse as $y_g(t,p T_s)$. Such signal has duration $\tau_{\rm tx}$ and bandwidth $B$, with time-bandwidth product $\tau_{\rm tx} B \gg 1$. It is first sampled at rate $\Delta t = 1/B_{\rm adc}$ (inverse of the sampling bandwidth of the receiver analog-to-digital converter $B_{\rm adc}$, which is usually larger than the effectively employed signal's bandwidth $B$) and converted to range dimension by $\Delta r = c\Delta t/2$, obtaining a discrete version of $y_g$, that is, $y_g(m \Delta r, pT_s)$, $m=1,..., M$, where $M$ is the number of range samples. Obviously, the quantity $M \Delta r$ corresponds to the radar maximum unambiguous range which is equal to $c\tau_{\rm rx}/2$.
The first processing step is to compute the cross-correlation between the Rx signal $y_g(m \Delta r, pT_s)$ and the Tx signal $s_g(m \Delta r, pT_s)$, obtaining the \textit{range-compressed} signal $x_g(m \Delta r, pT_s)$. Such range-compressed signal has effective range resolution equal to $c/2/B \geq \Delta r$, usually larger than the range sampling. 

%

\subsection{Doppler spectrum estimation} The second processing step is the estimation of the Doppler spectrum from $x_g(m\Delta r, pT_s)$, operating a windowing followed by a periodogram for each $m$-th range bin. The complex Doppler spectrum is defined as:  
\begin{equation}
\begin{split}
        X_g(m\Delta r, k \Delta f) & \hspace{-0.1cm}= \mathcal{F}_w\left\{x_g(m\Delta r, pT_s)\right\} \\
        & \hspace{-0.1cm} = \hspace{-0.1cm}\frac{T_s}{W} \hspace{-0.1cm}\sum_{p=0}^{N_p-1}\hspace{-0.1cm}x_g(m\Delta r, pT_s)\, w(p T_s) e^{- j 2 \pi k\Delta f p T_s }
\end{split}
\label{eq:amplitudeSectra}
\end{equation}
where $k=-N_p/2,....,N_p/2-1$ denotes the sample index in the Doppler frequency domain, and $\Delta f = 1/(N_p T_s)$ is the Doppler frequency resolution. 
The term $w(pT_s)$ denotes the complex weight of the window at the $p$-th pulse, aimed at reducing the sidelobes in the Doppler spectrum and ease the ground clutter filtering, and $W=\sqrt{\sum_{p=0}^{N_p-1} | w(pT_s)|^2}$ is a normalizing factor. A suitable choice is a Blackmann or Blackmann-Nuttal window. 
Note that the windowing operation is restricted to \(N_p < N_p'\) samples in order to obtain multiple Doppler instances of the same scene, which can subsequently be averaged (figure \ref{fig:DifferentialphaseDefinition}a,b).

The Doppler power spectral density is then calculated by the periodogram technique as: 
\begin{equation}
 S_g(m\Delta r, k \Delta f) = |X_g(m\Delta r, k \Delta f)|^2
 \label{eq:SpectralDensity}
\end{equation}
Then, the total Rx power on the $g$-th beam, $m$-th range bin is the integral of the power spectral density: 
\begin{equation}\label{eq:power}
    P_{\rm rx,g}(m \Delta r) =  \Delta f \sum_{k= -\frac{N_p}{2}}^{\frac{N_p}{2}-1} S_g(m\Delta r, k\Delta f).
\end{equation}
Eqs. \eqref{eq:SpectralDensity} and \eqref{eq:power} are the core of the whole processing that follows, and the key quantity from which the radar moments can be derived.

\subsection{Weather radar quantities}

For what follows, it is useful to recall the fundamental equations that govern the operation of a weather radar. Let us consider again a single beam, whose pointing angle is $(\theta_g,\phi_g)$. The total received power at range $r=m \Delta r$ is expressed as \cite{Bringi:2001}:
\begin{equation}
P_{\rm rx,g}(m\Delta r)=C_g \frac{Z_g(m\Delta r)}{(m\Delta r)^{2}}L_g^{2}(m\Delta r)
\label{eq:radar_eq}
\end{equation}
where: \textit{(i)} $(m\Delta r,\theta_g,\phi_g)$ identifies a resolution volume at distance $r=m\Delta r$ for the $g$-th beam \textit{(ii)} $C_g$ is the radar constant for beam $g$, which includes all the system parameters that pertain to the measurement system (e.g., emitted power, radiation pattern--explicitly dependent on the specific beam--, etc.) and need to be calibrated \textit{a-priori} in order to enable QPE (see Appendix \ref{AppendixA: radar equation} for further details), \textit{(iii)} $Z_g(m\Delta r)$ is the equivalent reflectivity factor measured in (mm$^6$ m$^{-3}$), \textit{(iv)} $L_g^{2}(m\Delta r)$ is the integral path loss due to atmospherics effects (eg. water vapor and rain) from the radar site up to distance $m\Delta r$.
\subsubsection{\textbf{Reflectivity factor}}
In \eqref{eq:radar_eq}, $Z_g(m\Delta r)$ depends by scattering and microphysical features of hydrometeors detected as follows:
\begin{equation}
Z_g(m\Delta r) \hspace{-0.1cm}=\hspace{-0.1cm}\frac{\lambda_0^4}{\pi^5 |K_w|^2}\hspace{-0.1cm}\int_{0}^{\infty} \hspace{-0.2cm}\sigma_b(D,\chi) N(D,m\Delta r, \theta_g,\phi_g)  dD.
\label{eq:Ze}
\end{equation}
%
 In the latter, $|K_w|^2$ is the dielectric factor and it is derived from the electric permittivity of water ($\approx$ 0.93 at S-C and X bands),  $\sigma_b(D,\chi)$ is the backscattering cross section (mm$^2$) of the drop with equivalent diameter $D$ illuminated with polarization $\chi$ and $N(D,r, \theta_g,\phi_g)$ is the drop size distribution in (m$^{-3}$mm$^{-1}$) at the resolution volume at position $(m\Delta r,\theta_g,\phi_g)$. 
 
\subsubsection{\textbf{Mean Doppler velocity}}
The mean Doppler velocity $V_{\rm D,g}$ at $m$-th range gate and $g$-th beam is defined as:
\begin{equation}\label{eq:MDV}
    V_{\rm D,g}(m \Delta r) = \frac{2}{\lambda_0} \underbrace{\frac{  \sum_{k= -\frac{N_p}{2}}^{\frac{N_p}{2}-1} k\Delta f\, S_g(m\Delta r, k\Delta f)}{P_{\rm rx,g}(m\Delta r)}}_{f_{\rm D,g}}
\end{equation}
in which factor $\frac{2}{\lambda_0}$ derives from the well known link between Doppler frequency and velocity: $f=-2\frac{v}{\lambda_0}$, where $\lambda_0=c/f_0$ is the carrier wavelength and $S_g$ and $P_{\rm rx,g}$ are defined in \eqref{eq:SpectralDensity} and \eqref{eq:radar_eq}, respectively .

\subsubsection{\textbf{Doppler spread}}
The Doppler spread ($W_{\rm D,g}$) quantifies the dispersion of Doppler shifts around the mean value and serves as an indicator of turbulence-induced effects:
\begin{equation}\label{eq:DVS}
    W_{\rm D,g}(m \Delta r) = \frac{2}{\lambda_0} 
    \sqrt{\frac{ \sum_{k= -\frac{N_p}{2}}^{\frac{N_p}{2}-1} (k \Delta f \hspace{-0.1cm}-\hspace{-0.1cm}f_{\rm D,g})^2 S_g(m\Delta r, k\Delta f)}{P_{\rm rx,g}(m\Delta r)}}.
\end{equation}

\subsubsection{\textbf{Minimum detectable reflectivity factor}}
One last quantity that needs to be introduced is the minimum detectable reflectivity (MDZ) which is strictly related to the radar system sensitivity (i.e. the expected minimum detectable signal in terms of $Z_g$). 
MDZ is obtained by inverting \eqref{eq:radar_eq} assuming $L_g^2=1$ and considering the power level from which the signal to noise ratio (SNR) is equal to 1. This corresponds to the minimum useful signal level that is theoretically detectable in noise at the receiver output. Consequently:
\begin{equation}
 \mathsf{MDZ}_g(m \Delta r)=\frac{P^{\min}_{\rm rx,g}}{C_g}(m \Delta r)^2
\label{eq:MDZ}
\end{equation}
In practice, the minimum power level yielding the MDZ is computed as $P^{\min}_{\rm rx,g}(m \Delta r) = P_{\rm noise} \,F_n \, \mathsf{SNR}_{\rm min}$, where $P_{\rm noise}$ is the thermal noise power, $F_n$ is the noise factor of the receiver and $\mathsf{SNR}_{\rm min}$ is the minimum SNR required by the user at the receiver output. Typically $\mathsf{SNR}_{\rm min}=1$ and values of $Z_g$ below MDZ are not detectable.

The main limitation to QPE by weather radars (and by BS-WRM as well) is the presence of ground static clutter, that biases the estimated precipitation and hinders QPE. In the following section we describe the employed processing chain for ground clutter filtering. To simplify the notation, we drop the beam index $g$ for simplicity, referring to single-beam processing, and we refer to the available datum as $x(m,p) \triangleq x_g(m \Delta r, pT_s)$, $X(m,k)\triangleq X_g(m \Delta r, k\Delta \nu)$, dropping the sampling intervals as well.

\subsection{Ground static clutter (GC) removal} \label{sec:GCremoval} Separate rain signal from the unwanted ground clutter (GC) is of fundamental importance to achieve reliable QPE results, especially if negative elevation angles are used, as with BS-WRM. Following literature (e.g. \cite{SPDPEO}), there can be two distinct approaches to suppress clutter in weather radar images:  
1) a two-step procedure, in which sufficiently long rain-free periods are collected to construct a static clutter map that facilitates clutter identification, followed by the application of a zero-Doppler notch filter to suppress the clutter contribution in precipitation-affected data \cite{Bringi:2001}; or 2) a single-step procedure that exploits the statistical and/or textural differences between clutter and precipitation, which are directly observable in the acquired data, thereby eliminating, or making less decisive, the need for a-priori acquisition of no-rain data.
Herein, due to the complexity in operating a dual system like BS-WRM, we follow an approach according to the second clutter suppression class although a refinement, which incorporates the collection of no-rain data, is also proposed.
The clutter suppression algorithm used in this work, is an improved extension of that proposed in \cite{SPDPEO}. The algorithm's block diagram is shown in figure \ref{fig:ClutterSuppressionBlockDiagram}. The starting point is the range compressed received signal for a given $g$-th beam: $x(m\Delta r,pT_s)$ (see section \ref{subsect:preprocessing} and figure \ref{fig:DifferentialphaseDefinition}a) which is processed following several steps:

\begin{figure}[!ht]
    \centering
     \includegraphics[width=1\columnwidth]{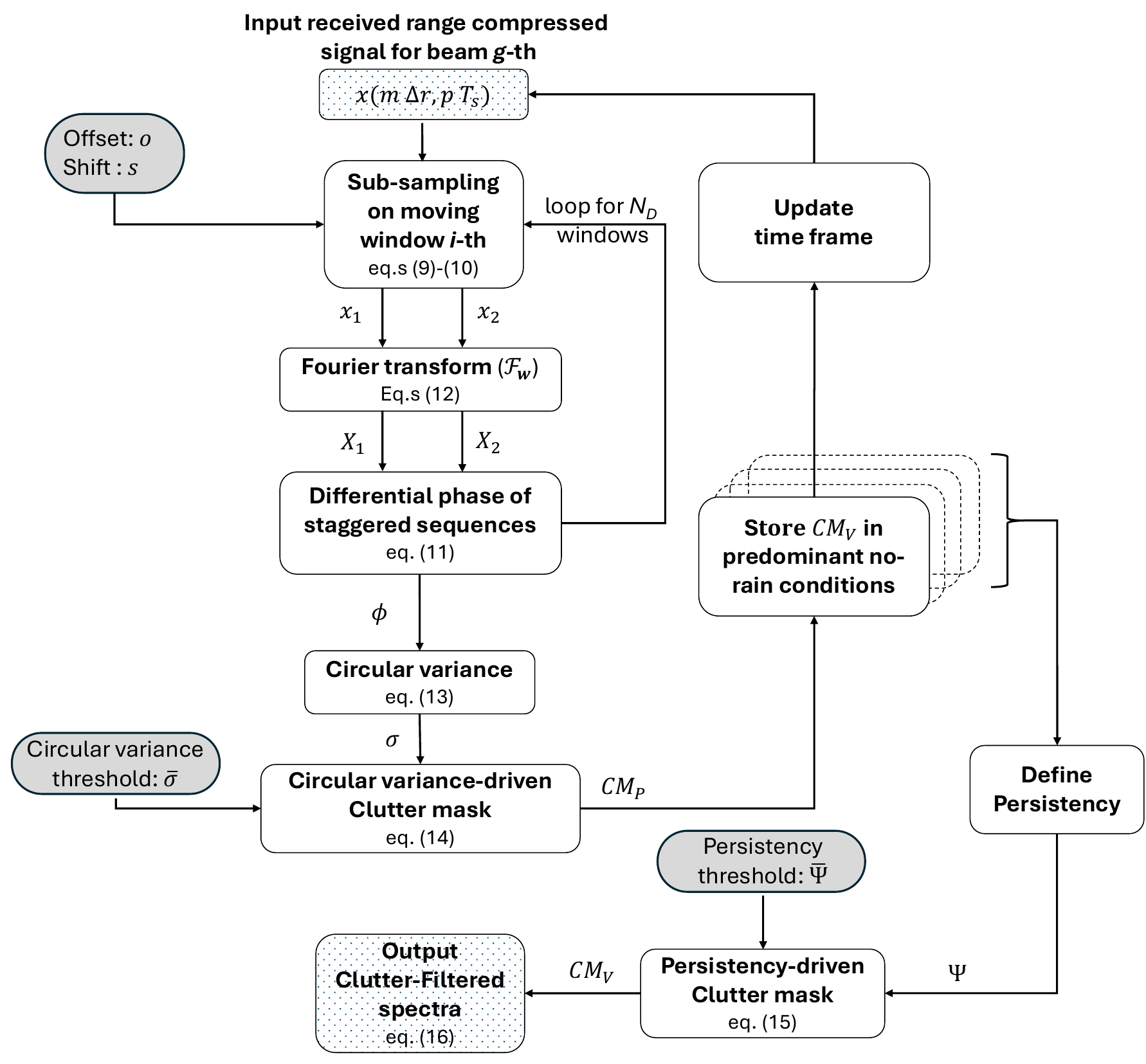}
  \caption{Block diagram of the clutter suppression algorithm used in the BS-WRM. Textured-filled boxes indicate input and output quantities, whereas those gray-shaded are the auxiliary input parameters.}
  \label{fig:ClutterSuppressionBlockDiagram}
  \end{figure}

\subsubsection{\textbf{Sub-sampling}}
The objective of the sub-sampling step is to construct two temporal staggered sequences, denoted by $x_1$ and $x_2$, from the original data  $x(m,p)$ such that, for each range bin $m$, the following condition holds:
    \begin{align}
        x_1(m, p;s,o) &= x(m, 2p-1 + s ), \label{eq:subsampling_x1} \\ 
        x_2(m, p;o,s) & = x(m, 2(p+o)+s) \label{eq:subsampling_x2} 
    \end{align} 
for $p=1,...,(N_p-o)/2$, being $N_p$ the number of samples in a given window at position $s$ (figure \ref{fig:DifferentialphaseDefinition}b,c,d).
The sequences $x_1$ and $x_2$ share a common shift $s = n \Delta s$  with $n$ an index from 0 to $N_D$-1 and $N_D$ the maximum number of windows that can be allocated into the original time sequence $x$ with a given shift gap ($\Delta s$). Within a generic window, $x_2$ is offset by a quantity $o$ with respect to $x_1$. By setting $s=0$ and $o=0$ we end up with even and odd sequences (compare figure \ref{fig:DifferentialphaseDefinition}d and e)) used in \cite{SPDPEO}.

\subsubsection{\textbf{Evaluate the differential phase (DP)}} 
\begin{figure}[!ht]
    \centering
     \includegraphics[width=1\columnwidth]{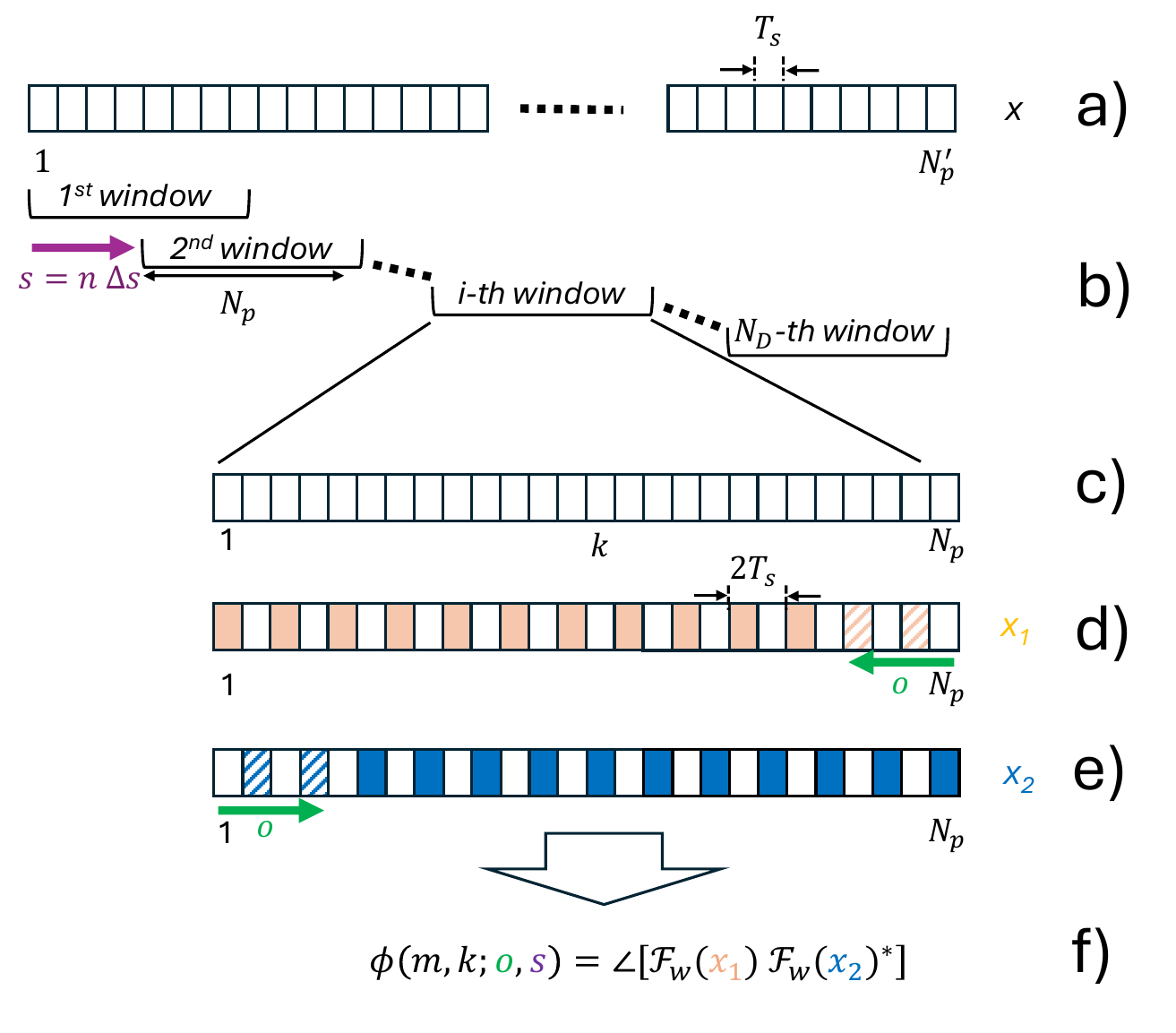}
  \caption{Representation of differential phase definition , $\phi(m,k;o,s)$ (f). The initial time sequence $x=x_g$ of $N_p^{'}$ samples for a single g-th beam (a) is then processed separately in $N_p$-wide moving windows  each of them spaced by a shift quantity $s$ (b). In a single window the time series of $N_p$ samples (c) is then divided in two distinct time series $x_1$ and $x_2$ as in eq. \eqref{eq:subsampling_x1} and \eqref{eq:subsampling_x2} of staggered samples (d), (e) orange and blue samples respectively, as a function of the offset $o$. Textured orange and blue samples in d) and e) are those discarded by the effects of the imposed offset.}
  \label{fig:DifferentialphaseDefinition}
  \end{figure}

the algorithm follows with the definition of the the differential phase in the spectral domain ($\phi$) as the Hermitian product of the two spectra of the sequences $x_1(m,p)$ and $x_2(m,p)$ (figure \ref{fig:DifferentialphaseDefinition}f),  as:
    \begin{equation}\label{eq:DP}
        \phi(m,k;o,s) = \angle \left(\,X_1(m, k;s,o) X_2^*(m, k;o, s)\right)
    \end{equation}
    where:
    \begin{align}
    \begin{split}
    X_1(m, k;s,o) &= \mathcal{F}_w\left\{x_1(m, p;s,o)\right\}\\
    X_2(m, k;s,o) &= \mathcal{F}_w\left\{x_2(m, p;s,o)\right\}
     \end{split}
    \end{align}
    denote the complex amplitude Doppler spectra of $x_1(m,p;s,o)$ and $x_2(m,p;o,s)$, respectively, through the $\mathcal{F}_w$ operator defined in \eqref{eq:amplitudeSectra}. The DP depends on the range–Doppler bin $(m,k)$ and on the offset–shift pair $(o,s)$. Note that the time series $x_1$ and $x_2$ have $(N_p-o)/2$ samples each and are sampled at $2T_s$ instead of $T_s$ (figure~\ref{fig:DifferentialphaseDefinition}(c–e)). This has the dual effects to reduce the spectral resolution $X_1$ and $X_2$ to  $\Delta f=1/[(N_p-o)T_s]$  with respect to $X$ in \eqref{eq:amplitudeSectra} that has $\Delta f=1/(N_p T_s)$ as well as, to halving the Doppler frequency unambiguous limits from $\pm \frac{1}{2T_s}$ to $\pm \frac{1}{4T_s}$, respectively. To restore the proper frequency resolution of $X_1$ and $X_2$ to that of $X$, a zero padding of $o/2$ samples on both $x_1$ and $x_2$ is applied. The final result is to have the $k$ index in \eqref{eq:DP} that varies from 1 to $N_p/2$ covering the spectral domain within $\pm \frac{1}{4T_s}$.

\subsubsection{\textbf{Calculate the circular variance (CV)}} 
\begin{figure}[!t]
    \centering
     \includegraphics[width=0.9\textwidth]{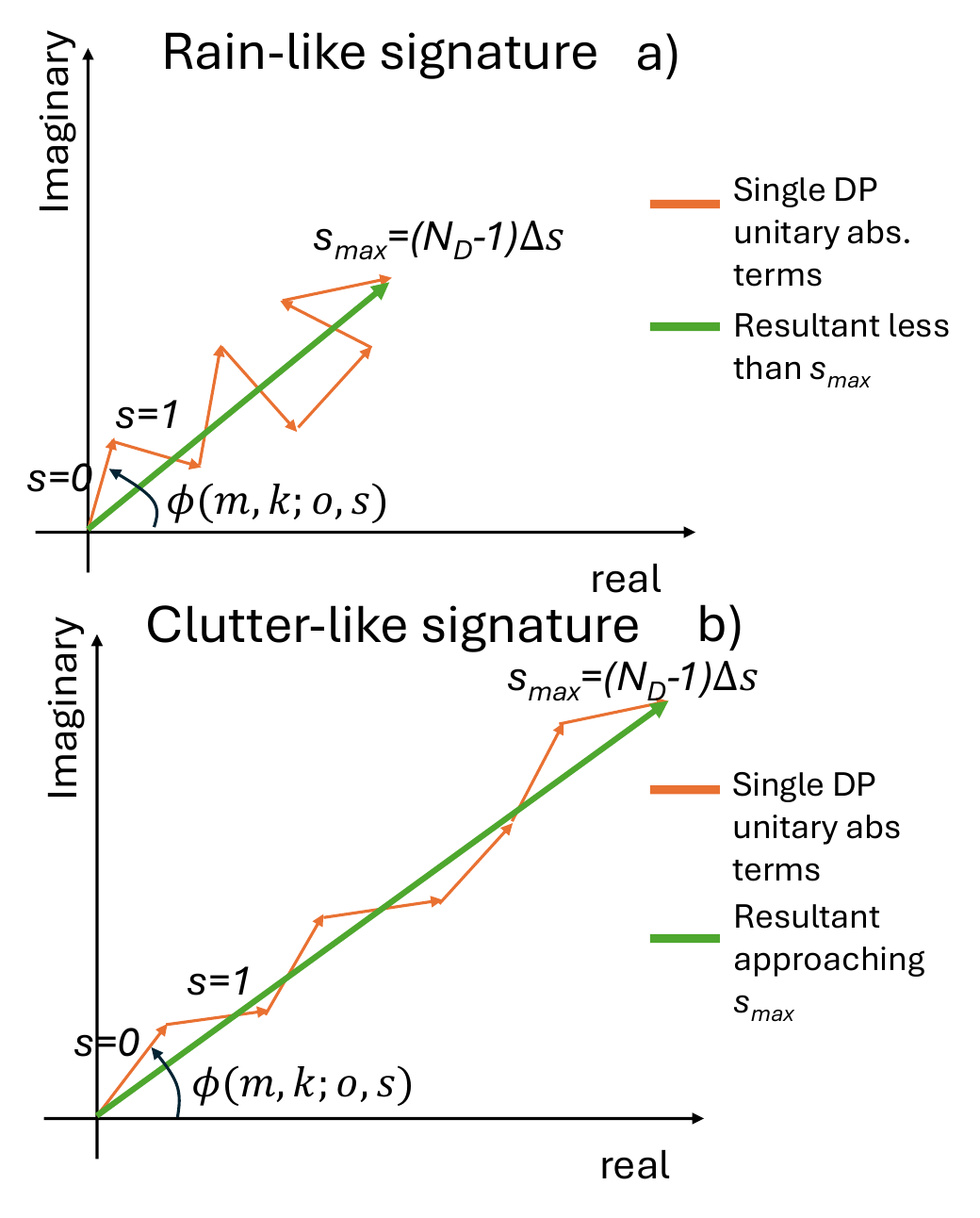}
     \caption{Schematization of CV principle. A case of unstable $\phi$ terms with $o=0$, producing a resultant much more less than unity and consequently $\sigma$ approaching one a); the complementary case in b).}
  \label{fig:CicrularVariance}
    \end{figure}

Since a single DP map $\phi(m,k;o,s)$ may be noisy due to the limited number of pulses used for its computation, we use the circular variance (CV) as a measure of the DP dispersion, to quantify its stability over different shifts ($s=0,...,(N_D-1)\Delta s$). The CV is defined as follows:
        \begin{equation}
        \sigma(m,k;o) = 1 - \bigg\lvert\frac{1}{N_D}\sum_{s = 0}^{(N_D-1)\Delta s} e^{j \phi(m,k;o,s)}\bigg\rvert^2
        \label{eq:circvariance}
    \end{equation}
    and it yields a more robust measurement of phase stability. When, $\sigma(m,k;o)$ approaches zero, phase at bin $(m,k)$ is stable indicating a stable target (i.e. likely clutter).
    Otherwise, it implies precipitation or noise. A conceptual representation of CV in \eqref{eq:circvariance}, is given in Figure \ref{fig:CicrularVariance} which allow to better understand the meaning of $\sigma(m,k;o)$. In particular, it can be observed that the individual exponential terms in \eqref{eq:circvariance}   (orange vectors in figure  \ref{fig:CicrularVariance})  may exhibit substantial noise depending on the observed scene, whereas the resulting mean component (green vector), which is strictly related to the definition of CV in \eqref{eq:circvariance}, still depends on the instability of the DP components, although it is more robust to their random fluctuations. In summary, $\sigma(m,k;o)$ constitutes a more stable measure of DP variability and is therefore well suited for clutter identification in a more homogeneous manner. However, $\sigma(m,k;o)$  is an explicit function of the relative offset between the two sequences, $o$, which is an external parameter to be tuned according to the specific precipitation event. Setting $o=0$, i.e., employing the method in \cite{SPDPEO}, may be inadequate for particular rain events characterized by reduced turbulence, namely a narrow Doppler spectrum, which in turn leads to an higher DP stability similar with that typically observed in the presence of ground clutter. Increase the offset $o>0$ should have the effect to rapidly reduce DP stability for rain signature while maintaining it higher for clutter. Therefore, assuming that clutter has a much higher phase stability than precipitation, the optimal value of $o$ would be inversely proportional to the Doppler spread of precipitation $W_{\rm D}$ in the original spectrum (and it would eventually depend on the range). 
    
\subsubsection{\textbf{Determine CV-driven clutter mask ($CM_V$)}} 
A clutter mask, denoted as $CM(m,k)$, is a binary map defined over the range–Doppler domain that characterizes the presence or absence of clutter for each range-Doppler bin $(m,k)$. Specifically, the function assumes the value $CM = 0$ in the presence of clutter and $CM = 1$ under clutter-free conditions. Starting from CV, the CV-driven  clutter map ($CM_V$) is defined as: 
    \begin{equation}
        CM_V(m,k;o) = \begin{dcases}
            0 & \sigma(m,k;o) \leq \overline{\sigma} \quad \text{for clutter}\\
            1 & \sigma(m,k;o) > \overline{\sigma} \quad \text{otherwise}
        \end{dcases}
        \label{eq:clutterMask1stGuess}
    \end{equation}
    where the threshold $\overline{\sigma}$ can be set either a-priori (e.g., $\overline{\sigma}=0.1$) or based on offline statistics of the clutter evaluated on a purposely collected dataset in which no-rain data dominate.
    It is worth to highlight that, differently from our procedure, the method in  \cite{SPDPEO} define $CM$ setting a threshold directly on the differential phase in \eqref{eq:DP} which is much more sensitive to noise than $\sigma$ in \eqref{eq:circvariance} (see figure \ref{fig:CicrularVariance} and related main text).
   Again, considering particular rain events characterized by reduced turbulence, by expressing the variations of CV as a function of the offset $o$ it can be shown that the latter becomes a key parameter to achieve an effective discrimination between clutter and rain. Increasing $o$ reduces phase coherency within those $(m,k)$-bins where rain signature is present, thus CV rapidly converges to unity while leaving that associated to clutter on lower values, as its phase dispersion is low even at larger lags $o$. As a consequence CV thresholding in \eqref{eq:clutterMask1stGuess} is expected to produce more accurate $CM_V$.     
   However, increments of $o$ produce a deterioration of Doppler resolution too, leaving also some overlap between the rain-clutter classes. Improvements of $CM_V$ can be achieved through the implementation of the next step.

\subsubsection{\textbf{Determine persistency-driven clutter mask ($CM_P$)}} 

    Since the clutter mask $CM_V$ in \eqref{eq:clutterMask1stGuess} is a real-time estimate that depends on the number of processed samples and/or spectra quality, the clutter filtering process may vary over time, with the obvious consequence of yielding a less homogeneous clutter suppression across the temporal dimension. To address this limitation, a refined clutter mask can be introduced, formulated on the basis of the concept of persistence. Persistency ($\Psi(m,k)$) can be defined as the number of occurrences for which $CM_V(m,k)$ in \eqref{eq:clutterMask1stGuess} equals 0 (i.e. indicating clutter) over a predefined temporal interval. This interval is selected such that no-rain conditions prevail, ensuring that the acquired scenes are mainly dominated by ground clutter.  
    Then, the persistency-driven refined clutter mask ($CM_P$)  can be retrieved as:
    \begin{equation}
        CM_P(m,k;o) = \begin{dcases}
            0 & \Psi(m,k;o) > \overline{\Psi} \quad \text{for clutter}\\
            1 & \Psi(m,k;o) \leq \overline{\Psi} \quad \text{otherwise}
        \end{dcases}
        \label{eq:clutterMaskPersistency}
    \end{equation}
    where $\overline{\Psi}$ is a persistency threshold that can be obtained similarly to the previous step by settling a quantile over the persistency map. It should be noted that the construction of \( CM_P \) requires the acquisition of multiple instances of \( CM_V \) so that its availability depends on the capability to collect data predominantly under non-rainy conditions, which is currently not a fully automated process. 
    
    Noticeably, for larger offsets, the rain-clutter distinction becomes increasingly pronounced, as an increase of the total acquisition time has the expected consequence of markedly reducing the rain signature, which by its nature exhibits higher temporal and spatial variability than clutter, the latter tending to persist in a more pronounced and stable manner.

\subsubsection{\textbf{Clutter filtered  spectra}}
    The clutter-filtered spectra $X_{ \rm filt}$ can be simply obtained by the following multiplication: 
    \begin{equation}
         X_{\rm filt}(m,k;o) = CM_\ell(m,k;o)\; X(m,k)
         \label{eq:clutterFilter}
    \end{equation}  
    and it is equal to zero in those pixels identified as clutter by $CM_\ell(m,k)=0$ where $\ell=V$ of $\ell=P$ to select \eqref{eq:clutterMask1stGuess} or  \eqref{eq:clutterMaskPersistency}, respectively. Note that, since $CM_\ell$ is constructed based on the definition of DP in \eqref{eq:DP}, which contains $N_p/2$ samples within the frequency interval $\pm 1/(4 T_s)$, whereas $X$ contains $N_p$ samples within the broader frequency interval $\pm 1/(2 T_s)$, the values of $CM_\ell$ are constrained to be equal to 1 outside the range $\pm 1/(4 T_s)$. Consequently, the product in \eqref{eq:clutterFilter} is to be interpreted as an element-wise (Hadamard) multiplication.

\subsection{Interpolation methodology}

The clutter-filtered power signal in range and Doppler $S_{\rm filt}(m,k) = |X_{\rm filt}(m,k)|^2$ needs interpolation in the missing clutter bins to restore the precipitation values and avoid biases in the QPE. Assuming that the Doppler spectrum of the precipitation at each range bin can be approximated by a Gaussian function \cite{Bringi:2001}, we can frame interpolation as the problem of fitting $S_{\rm filt}(m,k)$ with a Gaussian function, whose parameters are computed with usual iterative techniques. The final power spectrum will be 
\begin{equation}
    \widetilde{S}(m,k) = \begin{dcases}
        S_{\rm filt}(m,k) & \text{for } CM_\ell(m,k) = 1\\
        \widehat{A} e^{- \frac{(k- \widehat{\mu})^2}{2\widehat{\Sigma}^2}} & \text{for } CM_\ell(m,k) = 0
    \end{dcases}
    \label{eq:FiltPSD}
\end{equation}
and it is used  in \eqref{eq:power} and  \eqref{eq:Ze} for calculating the received power and the reflectivity factor, respectively,  and consequently  rain intensity retrieval as detailed in the next subsection. 

\subsection{Rain intensity retrieval}
\label{sect:RainIntensityRetrieval}

The retrieval of the rain intensity $R_g(m)$ at the $m$-th range gate and $g$-th pointing beam direction, starts by plugging the clutter-filtered and interpolated power spectral density, $\widetilde{S}(m,k)$ in \eqref{eq:FiltPSD}, into \eqref{eq:power}, yielding the total received power $\widetilde{P}_{\rm rx,g}(m)$, which includes the contribution from precipitation particles plus noise. Then, the equivalent reflectivity factor, $Z_g(m)$, is obtained by inverting the radar equation in \eqref{eq:radar_eq}. $R_g$ is finally obtained considering the widely used empirical power-law relation as follows: 
\begin{equation}\label{eq:aRb}
    Z_g(m\Delta r) = a R^b_g(m\Delta r)
\end{equation}
in which $Z_g$ and $R_g$ are in (mm$^6$m$^{-3}$) and (mm/h), respectively, whereas $a$ and $b$ are  the estimation coefficients. They are generally tuned by matching radar-measured reflectivity with rainfall accumulations recorded by rain gauges \cite{Bringi:2001}. However, due to variations of the drop size distribution  even over short spatial scales \cite{Tapiador:2010}, and differences in the sampling volumes of radar and rain gauge instruments, $a$ and $b$ exhibit substantial variability \cite{Battan:1973}. A quite consolidated \textit{a-priori} choice is to assume Marshall and Palmer coefficients ($a=200$, $b=1.6$). However, given the specific setting of BS-WRM with respect to custom weather radars in terms of polarization and working frequency, a new pair of coefficients: $a=92.0563$ and $b=2.1363$, are derived using a large set of 1.4 M of rainy minutes of measured $N(D)$ from the Italian Group of Disdrometry, (details in Appendix \ref{AppendixC:DisdrometerDatabase}). $N(D)$ from disdrometers are used into \eqref{eq:RainIntensityDef} and \eqref{eq:Ze}, together with electromagnetic simulations of the drop's radar cross section $\sigma_b$, to reproduce a statistic of $R_g$ and $Z_g$ in \eqref{eq:aRb} and tune coefficients $a$ and $b$, accordingly. It is worth nothing that the tuning process of $a$ and $b$ does not explicitly takes into account altitude displacement in $R_g$ and $Z_g$ since both these quantities are simulated at the same altitude as well as inhomogeneities in the radar resolution volume caused by non uniform beam filling effects.  Such arguments are indirectly considered by adding an uncertainty zero mean noise term on  $Z_g$  (1 dB error standard deviation in our case).
It should be borne in mind that for the conversion in \eqref{eq:aRb} to be effective, $Z_g$ has to be well calibrated. This means that the radar constant and attenuation factor in \eqref{eq:radar_eq_appendix} should be known with a sufficient degree of accuracy. On the two-way attenuation factor term, $L_g^2$, it is generally unknown unless the aid of polarimetry which is not used in the current setting of BS-WRM. However,  BS-WRM typically operates in a domain with a maximum range of 20 km  and consequently the $L_g^2$ term in \eqref{eq:pia} is not expected to increase considerably.

\section{Rain precipitation detection from base station: results from simulated scenarios}
\label{sect:ResultsSimScenarios}

\begin{figure}[!t]
    \centering
     \includegraphics[width=1.05\textwidth]{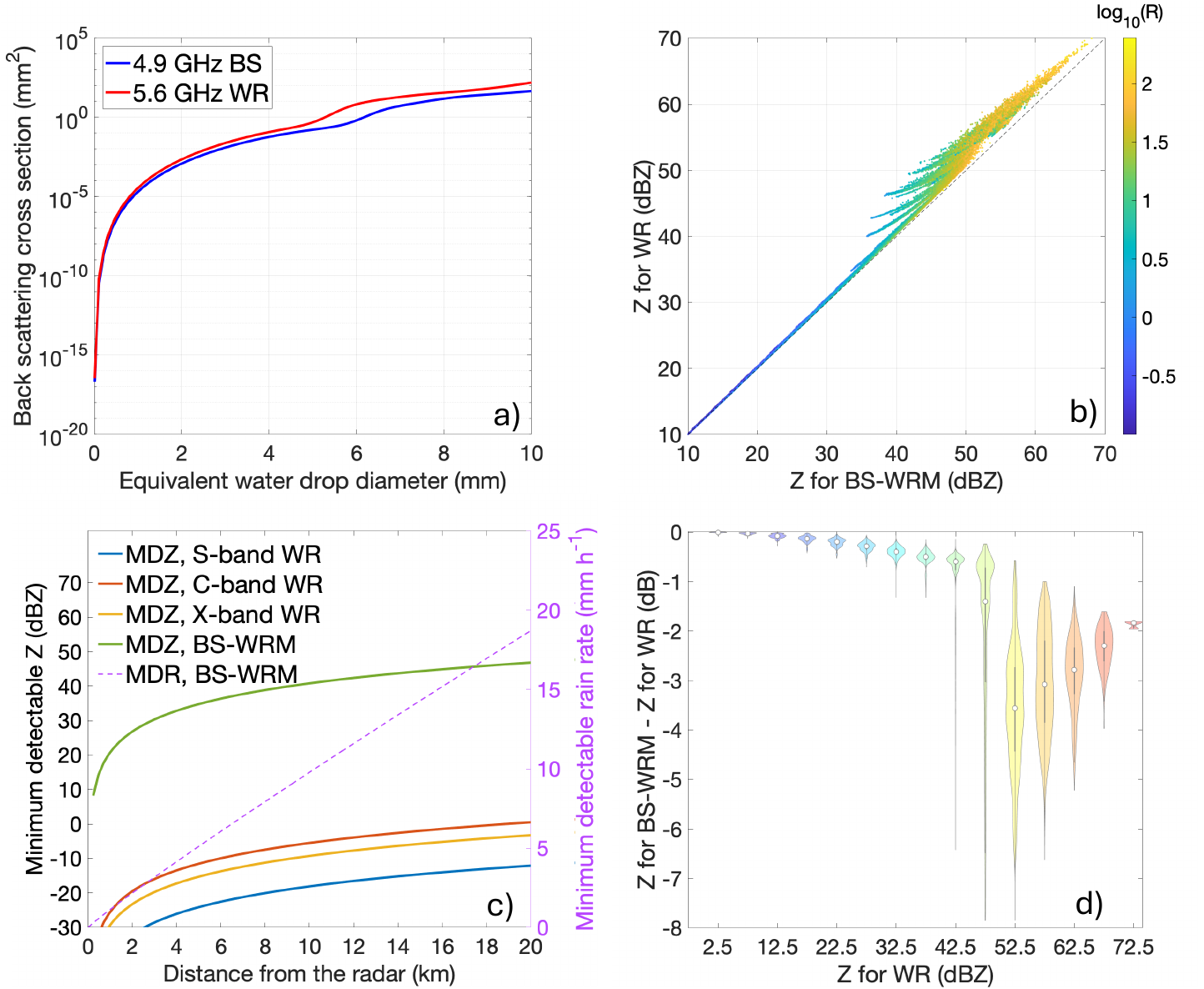}
  \caption{Simulations of copolar back scattering cross sections (a) for a base station in weather radar mode (BS-WRM) and  custom weather radar (WR). WR is assumed to work at 5.6 GHz with horizontal ($h$) polarization scheme whereas BS-WRM works  at  4.9 GHz having a 45$^\circ$ linear slanted polarization scheme for both transmission and reception. Comparisons in terms of reflectivity factor ($Z$) is in b); minimum detectable reflectivity ($MDZ$) for WR at various frequencies and BS-WRM is in c). For BS-WRM the minimum detectable rain rate  ($MDR=(\frac{10^{MDZ/10}}{a})^{1/b}$ with tuned coefficients  $a$=92.0563 and $b=2.1363$) is also shown (magenta dashed line). The $Z$-difference between BS-WRM and WR vs. $Z$ from WR, are in d).}
  \label{fig:radar_sim_Zcomparisons}
  \end{figure}
  
In this section the opportunity to use BS-WRM is proven from a simulation standpoint. 
A comparative analysis of BS-WRM rain detectability features with respect to those achievable from typical weather radar (WR) systems is presented. 
To this end, a simulation environment is built  by implementing eq. (\ref{eq:Ze}). The latter is run considering, $\sigma_b$ from EM simulations performed by a T-matrix code \cite{MISHCHENKO:1996}, whereas $N(D,r, \theta, \phi)$ is obtained by the GID (Gruppo Italiano di Disdrometria) ground disdrometer database \cite{Adirosi:2023} composed by 1.4M of rainy data (Appendix \ref{AppendixC:DisdrometerDatabase}).  In this case, $N(D,r, \theta, \phi)=N(D)$ since sparse point measurements from disdrometer alone do not allow to describe the spatial structure of $N(D)$.  In the simulation, for the BS-WRM we assumed 4.9 GHz carrier frequency with a +45$^\circ$ slanted linear polarization used for transmission and reception, whereas for WR we assumed a C-band radar (5.6GHz) with alternate orthogonal polarizations. For both systems, 0$^\circ$ elevation angle with respect the horizon is considered. The comparison of BS-WRM vs. WR is shown  in figure \ref{fig:radar_sim_Zcomparisons}a) in terms of radar cross sections, $\sigma_b$. \\

\begin{figure}[t!]
    \centering
     \includegraphics[width=1\columnwidth]{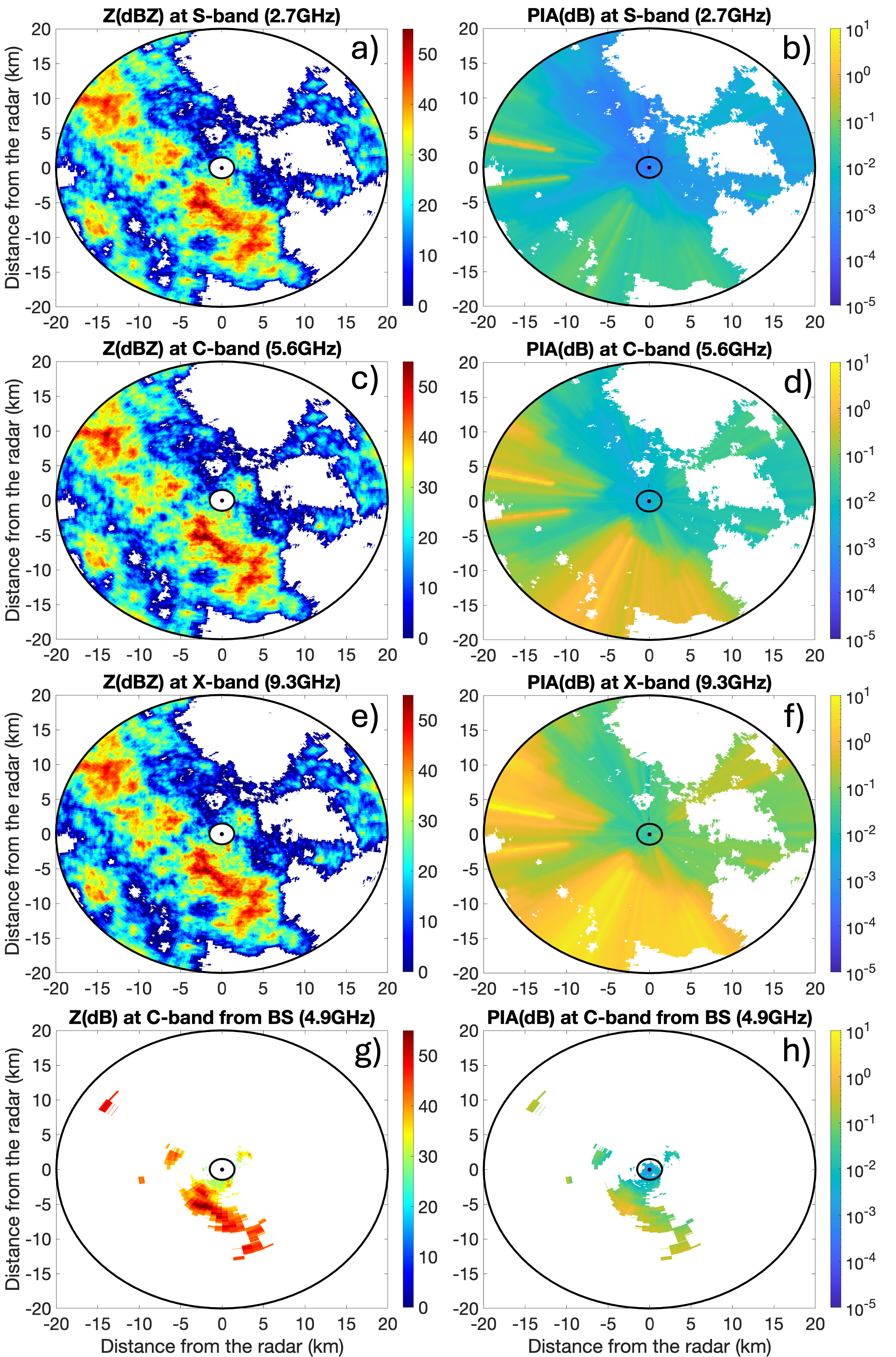}
  \caption{Simulations of radar reflectivity factor (dBZ) for typical S-, C-, X- weather radars (a,c,e) and BS  system (g) and corresponding two-way path integrated attenuation (d,d,f,h)}
  \label{fig:radar_sim}
\end{figure}

As it is clear there are differences in $\sigma_b$ which are more pronounced around 6 mm of equivalent water drop diameter with  $\sigma_b$ from BS-WRM constantly lower than that from WR. The same effect is visible in terms of $Z_e$ (\ref{fig:radar_sim_Zcomparisons}b) where the underestimation effects from BS-WRM compared to WR is more evident for $Z_e$ larger than 35dBZ. The Z domain  above 35dBZ will be likely sampled by the BS given the values on which the minimum detectable signal curve is based (\ref{fig:radar_sim_Zcomparisons}c). The same panel also highlight the difference in terms of sensitivity of BS with respect typical WRs as well as the minimum detectable rain rate (MDR) from BS which remains lower than 1 mm/h. The values of MDZ are obtained using the values listed in table \ref{tab:radar_specs}  (bold values) for WR,  whereas for BS-WRM working at $f_0$=4.9 GHz, a factor of the order of $10^{-3}$, $24$, $0.5$ and $18$ are applied to the bold values of $P_{\rm{tx}}$, $\tau_{\rm{tx}}$, $G_{\rm{tx}}$ and $\Delta \theta \cdot \Delta \phi$, referred to C-band in \ref{tab:radar_specs}, respectively, and applying them to eq.s \eqref{eq:radar_constant} and \eqref{eq:MDZ}.
Consequently, MDR is obtained by simply applying the $Z-R$ conversion (see section \ref{sect:RainIntensityRetrieval}) imposing Z= MDZ for the BS-WRM case. Finally, the Z difference between BS-WRM and WR, as a function of Z from WR (\ref{fig:radar_sim_Zcomparisons}d), evidences as for higher Z the BS-WRM  is expected to provide underestimates of the order 3.5 dB (on average) with peaks that can reach even 8 dB.
In summary, the analysis of figure \ref{fig:radar_sim_Zcomparisons} indicates as, although BS-WRM configuration is not optimal for rain observations, it is able to detect moderate to extreme events in an area within a maximum radius of 20 km from the BS site.
To visualize this concept, simulations are refined including antenna beam convolution. In this case, spatial DSD (i.e. $N(D,r, \theta, \phi)$ in eq. (\ref{eq:Ze}) are generated using a spatialization  procedure  \cite{Schleiss:2012} that is able to reproduce DSD features on a spatial domain, as those measured by ground disdrometers time series. To ingest antenna convolution in $Z_e$, $N(D,r, \theta, \phi)$ is generated at finer resolution than that of the radar system to be simulated. The simulations are conducted under the same assumptions employed for the generation of figure \ref{fig:radar_sim_Zcomparisons}. Simulations results are in figure \ref{fig:radar_sim}  which shows both $Z_e$ and 2-way Path integrated attenuation ($PIA$) and considers the appropriate $MDZ$ level (eq. \ref{eq:MDZ}) for each system.

Typical weather S-, C, X-band radars describe the full coverage domain similarly to each other (\ref{fig:radar_sim}a,c,e) due to their similar MDZ values. However, as expected,  a different impact in terms of 2-way PIA is noted  with increases of path losses as the frequency increases (b,d,f).
BS-WRM (g) suffer for limited sensitivity (i.e. higher MDZ as in figure \ref{fig:radar_sim_Zcomparisons}c) which implies that a large domain of the precipitation filed is invisible to the BS system with only medium to high $Z_e$ surviving. In this case, 2-way PIA (h) is not in general so relevant given the coverage limitations although in some extreme cases, care must be taken to address this issue adequately.

It is worth pointing out that the simulations produced do not take into accounts some important aspects like the ground clutter that is expected to be a dominant factor for BS-WRM since it likely operates in a highly antropicised context. 

\section{Discussion}
 Traditional remote sensing techniques for precipitation estimation can be often constrained by limited spatial and temporal resolution, as well as heterogeneous sampling and blind zones depending by the specific sensor considered. These limitations can become particularly pronounced in regions characterized by complex orography or dense urban environments, where near-surface retrievals are frequently degraded by environmental interference (eg. ground clutter). Conversely, in-site instruments such as rain gauges and disdrometers exhibit higher measurement accuracy but suffer from poor spatial representativeness.
Emerging opportunistic sensing approaches, including commercial and satellite microwave links, have offered a promising avenue to bridge the gap left by conventional remote sensing systems. However, the relatively low density of radio links per unit area combined with the absence of profiling capability, imposes significant constraints on the achievable spatial resolution of derived precipitation products.
 Given the widespread coverage of mobile networks, BS-WRM approach, is potentially a  high-resolution  gap filling solution especially in populated areas as urban contexts, where the capillarity of BS is particularly dense (figure \ref{fig:WR_and_BS_distributionWorldWide}a) compared to that of weather radars (figure \ref{fig:WR_and_BS_distributionWorldWide}b). 
The spatial density of meteorological radars and telecommunication BS exhibits pronounced differences, attributable to their distinct functional objectives. Surveillance weather radars, (e.g. NEXRAD in the U.S) are generally spaced several hundred kilometers apart (order of $\approx$200 $km$) to achieve extensive atmospheric surveillance. Assuming such a figure, and radar radius coverage  of $\approx$ 200$km$ too, the nominal density of a typical weather radar network is about one radar per 125600 $km^2$, or roughly 8 $\times 10^{-6}$ radars per $km^2$ \cite{Lynch_2023}. Within each single radar coverage there are blind zones of varying size that limit the rain detection closer to the ground. In contrast, due to the disparity related to the fundamentally different design objectives, telecommunication infrastructure is deployed at much higher densities \cite{chiaraviglio_2016}. Urban base station densities commonly reach 10 to 100 stations per $km^2$ (including 3G, 4G and 5G networks, with spatial increasing density expected for 6G). Rural base station densities are significantly lower, although still appreciable compared to the spatial extent of precipitation cores, falling in the range of 0.1 to 0.5 stations per $km^2$.  From these arguments  and the material presented in the previous sections, it is clear that BS-WRM can become a valid solution for gap filling current weather radar networks for intense or extreme rain events.
 In addition, the dual use of BS technology provides several notable advantages: (i) continuous 24/7 operational capability; (ii) very low data latency, as precipitation measurements are network-native ; (iii) unprecedented temporal and spatial sampling, enabled by relaxed coverage constraints, receiver bandwidth approximately an order of magnitude larger than that of conventional weather radars and the  multi-beam electronic steering. However, the BS approach also presents limitations: its relatively low sensitivity, mainly driven by low antenna gain and engaged power, restricts applicability to moderate-to-intense precipitation events only. Pervasive ground clutter fostered by near-horizontal orientation of antenna beams, although it can be effectively removed as described in Section \ref{sec:processing}, could partially block the BS signal contributing to further deteriorating the observed rain field. Path attenuation can further contribute to the signal losses increasing the blindness of the BS-WRM. However, as suggested in \cite{Lim:2011}, multiple overlapped observations from networked BS could help to alleviate such issue in the future. As a final consideration, it is worth to highlight as the expected amount of data that BS technology can deliver will be massive, opening to machine learning approaches of height resolution nowcasting of fast evolving precipitation extremes. Forecast from numerical weather prediction models will also benefit from BS data offering a floor for unconventional data assimilation at urban scales.

\section{Conclusions}
This study proposed an unprecedented methodology for rainfall observation, exploiting radio signals from BS, traditionally dedicated to mobile communications, as a completely new source of information. By applying a processing strategy analogous to that used in weather radar systems, this approach enables precipitation detection at significantly finer spatial and temporal scales than previously achievable.
With the aid of an EM simulation tool, it has been demonstrated that BS system, although it is not optimized for rain observation, is able track medium to high precipitation intensities within 20 km distance from each BS site. Special emphasis has been put ground clutter removal from Doppler spectrum acquired by BS which represent one of the main limitation of BS for weather applications. A clutter filtering technique, borrowed from those applied to weather radars, has been adapted to BS characteristics. The outcomes achieved clearly open to validation campaigns and massive use of BS in a networking way and vast areas.

\appendices
\section{Radar equation} 
\label{AppendixA: radar equation}

\numberwithin{equation}{section}
\setcounter{equation}{0}

In this appendix the key parameters driving the radar equation for distributed targets are discussed. The weather radar equation for the $g$-th beam \eqref{eq:radar_eq} is here reported for an easier reading:
\begin{equation}
P_{\rm rx}(r,\theta_g,\phi_g)=C_{g} \frac{Z(r,\theta_g,\phi_g)}{r^{2}}L^{2}(r,\theta_g,\phi_g)
\label{eq:radar_eq_appendix}
\end{equation}
in which ($r$, $\theta_g$, $\phi_g$) is the position of a radar resolution cell at distance $r$ from the radar for the $g$-th antenna beam boresight, whereas $P_{\rm rx}$ and $Z$ are the received average power and the equivalent reflectivity factor. In the following, without loss of generality, we simplify the notation by assuming a single beam and dropping the beam index $g$: $P_{\rm rx}(r)=P_{\rm rx,g}(r)=P_{\rm rx}(r,\theta_g,\phi_g)$, $C=C_g$, $Z(r)=Z_{g}(r)=Z(r,\theta_g,\phi_g)$, $L(r)=L_g(r)=L(r,\theta_g,\phi_g)$. The dependence on polarization $\chi$ is also dropped. 

\subsection{Path integrated  attenuation}
\label{sect:PIA}
The term $L^2$ is the two-way atmospheric path loss factor that is expressed as:
\begin{flalign}
& L^2(r)=\exp\Big[-\mathrm{2 \cdot PIA}(r)\Big]\nonumber\\
&=\exp\Bigg[-2\int_0^r \underbrace{\Bigg(\int_0^\infty \sigma_e(D, \chi) N(D) dD\Bigg)}_{\text{Specific attenuation: $k_e(r,\theta,\phi)$}} dr\Bigg].
\label{eq:pia}
\end{flalign}
$L$ depends on the extinction radar cross section of the drops ($\sigma_e$) (mm$^2$) and the drop size distribution $N(D)$ in (mm$^{-1}$m$^{-3}$), which in turn depends on the specific resolution volume, thus it is $N(D) \triangleq N(D,r,\theta_g,\phi_g)$. The integral along the diameter $D$ gives the specific attenuation ($k_e$) in (m$^{-1}$), while the integral over the range $r$ yields the one-way path integrated attenuation (PIA).


\subsection{Radar constant}

The radar constant ($C$) is divided into several terms, incorporating all the known system factors contributing to $P_{\rm rx}$. 
The expression of the radar constant is shown in \eqref{eq:radar_constant}, in which $10^{-18}$ is a units conversion term to have, in \eqref{eq:radar_eq_appendix} the power expressed in Watts when $Z$ is in (mm$^6 $m$^{-3}$), $r$ is in ($m$), $P_{\rm tx}$ is the transmitted peak power in ($W$), $G_{\rm max}$ is the one way antenna maximum power gain (in linear units), $\lambda_0$ is the carrier wavelength in ($m$), $|K_w|^2$ is related to the dielectric constant of water ($\approx$ 0.93 at S, C and X bands). The resolution volume is function of the HPBW along azimuth and elevation at boresight ($\phi=\phi_g,\theta=\theta_g$) $\Delta \phi$, $\Delta \theta$ and of the sampling interval along range (defined in Section \ref{subsect:preprocessing}) $\Delta r = c/(2 \cdot B_{\rm adc})$. 
The term $B \tau_{\rm tx}$ (with $B_{adc}>B$) is the product of the duration of the Tx waveform and its bandwidth and it represents the matched filter power gain after range compression at the receiver end. 
Since weather radar equation typically assumes pencil Gaussian beam  whose main lobe is characterized by standard deviations $\frac{\Delta \phi}{\sqrt{8 ln2}}$ and $\frac{\Delta \theta}{\sqrt{8 ln2}}$ along azimuth and elevation, respectively, an antenna correction factor ($F$) has to be introduced in $C$ to account for the non-Gaussian pattern of the specific antenna used by the radar/BS (i.e. the antenna array of the BS).
%
\begin{figure*}[ht!]
    \begin{equation}\label{eq:radar_constant}
        C = 10^{-18}\cdot \underbrace{\frac{P_{\rm tx} G_{\rm max}^{2} \lambda_0^{2}}{(4\pi)^{3}}}_{\begin{array}{c} \scriptsize {\text{Trasmission}}\\ \scriptsize {\text{term}} \end{array}} \cdot \underbrace{\frac{\pi^{5} |K_w|^2}{\lambda_0^{4}}}_{\begin{array}{c} \scriptsize {\text{Rayleigh}}\\ \scriptsize {\text{scattering}} \end{array}} \cdot  \underbrace{\frac{\pi \Delta \phi \Delta \theta}{8 \log 2 } \Delta r}_{\begin{array}{c} \scriptsize {\text{Resolution}}\\ \scriptsize {\text{volume}} \end{array}} \cdot \underbrace{F}_{\begin{array}{c} \scriptsize{\text{Antenna}}\\\scriptsize {\text{correction factor}} \end{array}}\cdot \underbrace{B \tau_{\rm tx}}_{\begin{array}{c} \scriptsize {\text{Range}}\\ \scriptsize {\text{compression gain}} \end{array}} 
    \end{equation}
    \hrulefill
\end{figure*}
\begin{equation}\label{eq:antenna_factor}
    F = \frac{8 \log 2}{\pi \Delta \phi \Delta \theta}\int_{-\pi/2}^{\pi/2} \int_{-\pi/2}^{\pi/2} f^2(\phi-\phi_g,\theta-\theta_g) \cos \phi \,d\phi \,d \theta
\end{equation}
In the latter, $f$ is the  \textit{normalized} power radiation pattern of the radar/BS antenna so that its maximum value is $f(\phi_g,\theta_g)=1$.   
The square factor in \eqref{eq:antenna_factor} is due to the monostatic radar configuration (i.e. same antenna for transmission and reception). 
It is important to remark that $F$, as well as,  the maximum antenna power gain, $G_{\rm max}$,  are generally beam specific, as the radiation pattern $f(\phi-\phi_g,\theta-\theta_g)$ changes as a function of the $g$-th beam.

\section{Rain intensity definition}    
\label{AppendixB:RainIntensity}

Rain intensity at altitude $h$ above the sea level (asl.) is defined as a statistical moment of the drop size distribution ($N(D)$) as follows:
\begin{equation}
R(h)=6\pi10^{-4} \int_{D_{min}}^{D_{max}} v(D,h)  N(D)  D^3 \,  dD 
\label{eq:RainIntensityDef}
\end{equation}
where $v(D,h)$ is the terminal fall velocity in still air of a liquid drop of equivalent size $D$ at altitude $h$ asl. $R$ in \eqref{eq:RainIntensityDef} is in (mm/h) when $D$ is in (mm) and $v$ is in (m/s). The terminal drop's velocity can be modeled as in \cite{Atlas:1973} with modification to take into account the altitude $h$ \cite{Porcu:2014,Thurai_2005}:
\begin{equation}
v(D,h)=\underbrace{(9.65-10.3e^{-0.6D})}_{v_0(D)} \bigg(\frac{\rho_0}{\rho_h}\bigg)^{(0.375+0.025D)} 
\label{eq:TerminalVel}
\end{equation}
In \eqref{eq:TerminalVel}, $v_0$ is the  terminal drop's fall speed at sea level whereas $\rho_h$ and  $\rho_0$, both in (kg/m$^{-3}$) are the air density at sea and $h$ level, respectively.  
$R(h)$ in \eqref{eq:RainIntensityDef} can be easily modeled knowing the $N(D)$ term  which is typically derived by disdrometers and the air density profile which can be derived by radiosoundings (i.e. temperature, pressure and humidity sensors on balloons) or considering the international standard atmosphere model as follows: 
\begin{equation}
\rho_h=\rho_0 (1-\alpha h)^\beta 
\label{eq:AirDensityProf}
\end{equation}
with $\rho_0=1.225$ ($kg \cdot m^{-3}$), $\alpha=2.2558\cdot 10^{-5}$ and $\beta=4.256$. 

\section{Disdrometer database}    
\label{AppendixC:DisdrometerDatabase}
Disdrometer are non-captative instruments able to quantify the particle size distribution (i.e. number of hydrometeros, $N$ corresponding to size equivalent diameter $D$) passing through the sensing area of the instrument in real time. Due to the ability in providing $N(D)$, which drives $Z$ and $R$ in eq.s, \eqref{eq:Ze} and \eqref{eq:RainIntensityDef}, disdrometers are more and more often a key elements to built \textit{ad-hoc} $Z-R$ relationships as done in section \ref{sect:RainIntensityRetrieval}.

The disdrometers used are those collected in Italy by the Italian Group of Disdrometry (GID) network  (\url{https://www.gid-net.it/}). GID was borne in 2021 thanks to a spontaneous collaboration of different Italian institutions (including research centers, universities, and environmental regional agencies) that manage disdrometers over the Italian peninsula. The GID network coordinates field campaigns, data sharing, and intercomparison activities across Italy, aiming to improve the accuracy and consistency of rainfall measurements.
Nowadays the GID networks consists of 27 disdrometers distributed across Italy, although due to technical issues it is possible that data from some devices are not available continuously. All the disdrometers of the GID network are laser disdrometer, most of them are Laser Precipitation Monitoring (LPM) of Thies Clima GmbH, however there are also some Parsivel2 of OTT GmbH. The longest time series of DSD data was the one collected in Rome that consists of more than 13 years of data (i.e. from September 2012 to December 2025). There are 3 disdrometers located at high altitude (i.e. more than 1000 m above sea level) that can likely collect snow or solid hydrometeors although this is not of interest for the present work. All disdrometers in the GID network use a uniform data-processing workflow so that their output (i.e time series of 1-minute DSDs) is provided in a standardized format. This harmonization is key to enabling cross-site comparisons and aggregated analyses. More information on the processing of the raw data adopted by GID to retrieve quality control DSDs is reported in \cite{Adirosi:2023}. 
The GID database is made freely available under a CC BY 4.0 licence by Zenodo ad it is updated yearly (the last  version used in this study is updated to 2024 and it is available here \cite{Adirosi:2025}). Such a version consists of  DSDs of almost 1.4 million rainy minutes.

\bibliographystyle{IEEEtran}
\bibliography{biblio}







\end{document}